\documentclass{elsart}
\usepackage{graphics}
\usepackage{graphicx}
\usepackage{epstopdf}
\usepackage{epsfig}
\usepackage{amssymb}
\usepackage{makeidx}
\usepackage{amsfonts}
\usepackage{amstext}
\usepackage{amsmath}
\usepackage{amsbsy}
\usepackage{wasysym}
\usepackage[vflt]{floatflt}

\def\mip{\it{mip}\rm}
\def\WArP{\sf{WArP}\rm~}
\def\N2{N$_2$~}
\def\A{\kern+.6ex\lower.42ex\hbox{$\scriptstyle \iota$}\kern-1.20ex a}
\def\E{\kern+.5ex\lower.42ex\hbox{$\scriptstyle \iota$}\kern-1.10ex e}

\makeindex
\begin{document}

\begin{frontmatter}

\title{Effects of Nitrogen contamination\\ 
 in liquid Argon}
\vspace*{-0.6cm}
{\large\sf WArP Collaboration}\\
{\small
\author[UAquila]{\small R.Acciarri}{,}
\author[LNGS,UAquila]{\small M.Antonello}{,}
\author[Padova]{\small B. Baibussinov}{,}
\author[UPadova]{\small M.Baldo-Ceolin}{,} 
\author[UPavia]{\small P.Benetti}{,}
\author[Princeton]{\small F.Calaprice}{,} 
\author[Pavia]{\small E.Calligarich}{,} 
\author[UPavia]{\small M.Cambiaghi}{,}
\author[UAquila]{\small N.Canci}{,}
\author[UNapoli]{\small F.Carbonara}{,} 
\author[UAquila]{\small F.Cavanna\corauthref{cor}}{,}
\corauth[cor]{Corresponding authors}
\author[UPadova]{\small S. Centro}{,} 
\author[Napoli]{\small A.G.Cocco}{,}
\author[UAquila,LNGS]{\small F.Di Pompeo}{,}
\author[UNapoli]{\small G.Fiorillo}{,} 
\author[Princeton]{\small C.Galbiati}{,}
\author[Napoli]{\small V. Gallo}{,}
\author[LNGS,UAquila]{\small L.Grandi}{,}
\author[Padova]{\small G. Meng}{,}
\author[UAquila]{\small I.Modena}{,}
\author[Pavia]{\small C.Montanari}{,} 
\author[LNGS]{\small O.Palamara}{,} 
\author[LNGS]{\small L.Pandola}{,}
\author[Padova]{\small F. Pietropaolo}{,} 
\author[Pavia]{\small G.L.Raselli}{,} 
\author[Pavia]{\small M.Roncadelli}{,}
\author[Pavia]{\small M.Rossella}{,} 
\author[LNGS]{\small C.Rubbia}{,}
\author[LNGS]{\small E.Segreto\corauthref{cor}}{,}
\author[Cracow,UAquila]{\small A.M.Szelc}{,}
\author[Padova]{\small S. Ventura}{,}
\author[Pavia]{\small C.Vignoli}
}
\address[UAquila]{Universit\`a dell'Aquila e INFN, L'Aquila, Italy}
\address[LNGS] {INFN - Laboratori Nazionali del Gran Sasso, Assergi, Italy}
\address[Padova]{INFN - Sezione di Padova, Padova, Italy}
\address[UPadova]{Universit\'a di Padova e INFN, Padova, Italy}
\address[UPavia]{Universit\'a di Pavia e INFN, Pavia, Italy}
\address[Pavia]{INFN - Sezione di Pavia, Pavia, Italy}
\address[Princeton]{Princeton University - Princeton, New Jersey, USA}
\address[UNapoli]{INFN - Sezione di Napoli, Napoli, Italy}
\address[Napoli]{Universit\'a di Napoli e INFN, Napoli, Italy}
\address[Cracow]{IFJ PAN, Krakow, Poland}
\vspace*{-0.7cm}
\begin{abstract}
A dedicated  test of the effects of Nitrogen contamination in liquid Argon has been performed at the 
INFN-Gran Sasso Laboratory (LNGS, Italy) within the \WArP R\&D program. \\
A detector has been designed and assembled for this specific task
and connected to a system for the injection of controlled amounts of gaseous Nitrogen into the liquid Argon. 
Purpose of the test is to detect the reduction of the Ar scintillation light emission as a function of the amount of the Nitrogen contaminant injected in the Argon volume. 
A wide concentration range, spanning from $\sim$10$^{-1}$ ppm up to $\sim$10$^{3}$ ppm, has been explored.\\
Measurements have been done with electrons in the energy range of minimum ionizing particles ($\gamma$-conversion from radioactive sources).
Source spectra at different Nitrogen contaminations are analyzed, showing sensitive 
reduction of the  scintillation yield at increasing concentrations. \\
The rate constant of the light quenching process induced by Nitrogen in liquid Ar has been found to be $k$(N$_2$)=0.11 $\mu$s$^{-1}$ppm$^{-1}$.\\
Direct PMT signals acquisition at  high time resolution by  fast Waveform recording allowed to extract with high precision the main characteristics of the scintillation light emission in pure and contaminated LAr. In particular, the decreasing behavior in lifetime and relative amplitude of the slow component is found to be appreciable from $\mathcal{O}$(1 ppm) of Nitrogen concentrations.     
\end{abstract}

 \end{frontmatter}

\tableofcontents

\section{Introduction}
\label{sec:Introd}
Dark Matter search in the form of WIMPs is of primary interest in the present astroparticle physics scenario. Direct detection of Dark Matter 
with noble gases liquified as target medium is one of the most promising line of development in experimental technology. 
Argon is an ideal medium and the feasibility of Ar-based detectors has been firmly proved by the R\&D study of the \WArP Collaboration \cite{warp2}. The two-phase (liquid-gas) technology developed by \WArP is based on the simultaneous detection of both signals produced by ionization events in liquid Argon (LAr): free electron charge and scintillation light.\\
In the case of the ionization charge, the main limitation to its full collection comes from electron attachment process due to contaminations of electro-negative impurities like Oxygen present at residual level in (commercial) Argon. Detailed studies have been performed and led, since the pioneering work of the Icarus Collaboration \cite{icarus}, to the development of adequate purification systems (Oxygen reactants and molecular sieves, down to $\le 0.1$ ppb of Oxygen equivalent concentration), normally employed in experimental applications. \\
On the other hand, the effects of impurities on the scintillation light yield are rather less precisely explored. 
Quenching (i.e. non-radiative) process in two-body collision of impurity-molecules with Ar$^*_2$ excimer states, otherwise decaying with VUV light emission, may take place depending on the type of impurity and the concentration level. 
This led to include within the \WArP R\&D program investigations on Oxygen and Nitrogen contamination effects on LAr scintillation light. Two experimental tests have been thus performed. \\
The case of Nitrogen (N$_2$), present in commercial Argon at the level of 1 to 10 ppm, depending on the selection grade, is the subject of the present study (while results from the test with Oxygen have been reported separately \cite{O2_test}). \\
Quenching of the light yield in N$_2$-contaminated LAr has been found, in particular of the slow component of the scintillation emission. This effect is relevant not only because it imposes some  limitations on the full collection of the light available, but also because it deteriorates the detector capability of Pulse Shape discrimination of background events (electrons from $\gamma$ interactions) from the signal (Ar-recoils, potentially induced by WIMP interactions) \cite{warp2}.

\section{Features of the scintillation radiation in LAr}
\label{sec:scint_light}
It is experimentally established and theoretically well understood that 
recombination and de--excitation processes following the passage of ionizing particles in liquid Argon (LAr) finally originate scintillation radiation \cite{kubota1}, \cite{doke1}. In particular, interactions of ionizing particles in LAr 
cause the formation of both electron-hole (Ar$^+$) pairs and Ar$^*$ excited atoms (the ratio Ar$^*$/Ar$^+$
being 0.21~\cite{doke2}). Ar$^*$ excited atoms lead to 
the formation of the Ar$^*_2$ low excited dimer through collision with Ar atoms (``self-trapping'' process). On the other hand, 
Ar$^+$ ions also lead to the formation of Ar$^*_2$ through a number of subsequent processes including 
recombination~\cite{kubota2}. The excited dimer states formed 
in both cases in LAr are recognized to be the singlet $^1\Sigma_u$ and the triplet 
$^3\Sigma_u$ excimer states in the {\it M}-band, characteristic of the solid Argon structure \cite{koch}. The 
rise-time corresponding to excimer formation and relaxation is very fast, in the sub-nanosecond range 
for both components. \\
The de-excitation processes to the dissociative ground state $^1\Sigma_g$ lead to scintillation 
light emission in the Vacuum Ultra-Violet (VUV) region:
\begin{equation}
Ar^*_2 \rightarrow 2~Ar + 1~\gamma
\label{eq:Ar_scint}
\end{equation}
In liquid phase, various electron-to-ion recombination mechanisms along the ionization track are active, 
depending on the type of the ionizing particle and on its LET (Linear Energy Transfer, i.e. the 
specific energy loss along the path) \cite{doke2}. These mechanisms affect the number of excited dimers 
Ar$^*_2$ produced per unit of deposited energy, as well as the relative populations of the singlet 
and triplet states.
For a minimum ionizing particle (\mip) the photon yield has been measured to be 
$4.0\times 10^4 ~\gamma$/MeV \cite{doke2}.\\
The $\gamma$ decay spectrum of both excimer states has been extensively investigated.
It is well represented by a gaussian 
shape, peaking at $\lambda$ = 127 nm with $FWHM \simeq 6$ nm (at boiling point T=87.3 K, $\rho=1.395$ g/cm$^3$) \cite{morikawa}.
The time dependence of the photon emission is known less precisely. The various measured values of both 
the excimer lifetimes and of the relative amplitudes reported in literature are in fact quite different 
with each other. To a first approximation, in all measurements the overall scintillation light emission 
exhibits a double 
exponential decay form (see \cite{morikawa} for a complete compilation of available data), characterized by 
two very different components: a {\it fast} component, with a time constant $\tau_S$ in the 2 ns to 6 ns 
range, and a {\it slow} component, with a time constant $\tau_T$ in the 1100 ns  to 1600 ns range. 
These are associated to the lifetimes of the singlet $^1\Sigma_u$ and of the triplet $^3\Sigma_u$ states 
in LAr, respectively. \\
In addition to these fast and slow components, an {\it intermediate} component (with decay time around 40 ns) 
has been also sometime reported in literature \cite{hitachi} \cite{himi}, the origin of which was never investigated in details 
nor definitively confirmed. 

While the time constants do not depend appreciably on the ionization density, the amplitude 
ratio $A_S/A_T$ of the singlet and triplet states is found to be strongly affected by the ionization density  \cite{kubota3}. 
In particular, all measurements show an enhancement of the $^1\Sigma_u$ formation ({\it fast} component) 
at higher deposited energy density\footnote{This effect is contrary to the 
effect in organic scintillators, in which the relative intensity of the slow component increases with 
increasing specific ionization density.}. 
As an example, the relative amplitude for the fast and for the slow component  in case of a 
\mip~ is reported to be $A_S/A_T=0.3$~\cite{hitachi} (i.e. $A_S = 23$\% and  $A_T = 77$\% respectively), while for heavily ionizing particles the intensity ratio increases (e.g. $A_S/A_T=1.3$ for $\alpha$-particles and =3 for nuclear recoils, but higher values are given elsewhere \cite{carvalho}).  This wide separation is an important feature of the scintillation signals in LAr, leading to define robust Pulse Shape Discrimination 
criteria suitable for particle identification.

The lack of accuracy/consistency in the presently available data can presumably be ascribed to (i) light quenching 
from residual concentration of impurities diluted in LAr, (ii) limited time resolution employed in past 
experiments, (iii) dependence on the LAr density (not systematically reported or accounted for). \\
  
\subsection{Light quenching from diluted N$_2$ impurities in LAr}
\label{sec:light_quenching}  
The de-excitation process rate can be described by a first-order rate law, characterized by two
decay time constants $\tau_S$ and $\tau_T$ associated to the singlet $^1\Sigma_u$ and the 
triplet $^3\Sigma_u$ excimer states of Ar$^*_2$, respectively ($j=S,T$):
\begin{equation}
\frac{d~[Ar^*_2]_j}{dt}~=~-~\frac{1}{\tau_j}~[Ar^*_2]_j~~~\Rightarrow~~~[Ar^*_2]_j(t)~=~[Ar^*_2]_j(0)~e^{-t/\tau_j}
\label{eq:1st_ord_law_VUV}
\end{equation}
Assuming that one scintillation photon per Ar$^*_2$ dissociation is produced, 
the time dependence of the scintillation light emission in 
{\it pure} LAr can thus be represented by the following probability distribution function (pdf):
\begin{equation}
\ell(t)~=~\frac{A_S}{\tau_S}~\exp(-\frac{t}{\tau_S})~+~\frac{A_T}{\tau_T}~\exp(-\frac{t}{\tau_T})
\label{eq:light_pdf}
\end{equation}
with the sum of the relative amplitudes constrained to unity ($A_S+A_T=1$) so that
\begin{equation}
\int_0^\infty \ell(t) dt~=~(A_S~+A_T)~=~1.
\label{eq:norma}
\end{equation}
Light quenching due to (residual and unknown) impurity concentration in LAr is often indicated as the dominant 
source of uncertainty in the available data on LAr intrinsic properties.\\
Residual concentration at the ppm (part per million) level of N$_2$, O$_2$, H$_2$O and CO+CO$_2$ contaminants 
(whose presence is usually indicated in commercially available Argon) can lead to a substantial reduction of the 
scintillation light intensity,  mainly due to the decrease of the {\it slow} component amplitude.\\
Argon purification systems (Oxygen reactants and molecular sieves) are known to be very effective in reducing 
the O$_2$ and H$_2$O, CO+CO$_2$ contamination up to a negligible level ($\le 0.1$ ppb, part per billion). At present, methods 
for removing the N$_2$ residual content in Argon are instead less commonly used in experimental applications\footnote{O$_2$ and CO+CO$_2$ are known to be electro-negative molecules. Their high $e$-affinity induces ionization electron attachment, with dramatic effects on detectors based on charge collection (LAr-TPC). N$_2$ molecules are not electro-negative \cite{barabash}, and therefore no appreciable effects on the free electron charge can be expected.}, and the effects 
of  N$_{2}$ contamination are rather poorly known.\\
The main reaction to be considered
is the quenching process in two-body collision of N$_2$ impurities with Ar$^*_2$ excimer states \cite{himi}:  
\begin{equation}
Ar^*_2~+~N_2~\rightarrow~ 2 Ar~+~N_2
\label{eq:quench}
\end{equation}
This non-radiative collisional reaction is in competition with the the de-excitation process leading to VUV light 
emission. As a result, a  sensitive quenching of the scintillation light yield is expected, in particular for 
lightly ionizing particles.
In fact, the quenching process leads to the decrease of the excimer concentration [Ar$^*_2$], while the 
contaminant concentration  [N$_2$] stays constant in time. To a first approximation, also for this case a 
first-order rate law can be assumed and characterized by the rate constant $k$:
\begin{equation}
\frac{d~[Ar^*_2]}{dt}~=~-~k~[N_2]~[Ar^*_2] ~~~\Rightarrow~~~[Ar^*_2](t)~=~[Ar^*_2](0)~e^{-k[N_2]~t}
\label{eq:1st_ord_law}
\end{equation}
The value reported in literature is  $k\simeq$ 3.8$\times 10^{-12}$ cm$^3$ s$^{-1}$ \cite{himi}
(equivalent to 0.11~ppm$^{-1}$ $\mu$s$^{-1}$).

The time dependence of the scintillation light emission in N$_2$-contaminated LAr can be represented by an 
expression formally identical to Eq.\ref{eq:light_pdf}:
\begin{equation}
\ell'(t)~=~\frac{A'_{S}}{\tau'_{S}}~\exp(-\frac{t}{\tau'_{S}})~+~\frac{A'_{T}}{\tau'_{T}}~\exp(-\frac{t}{\tau'_{T}})
\label{eq:quench_pdf}
\end{equation}
where $\tau'_{j}$ [$j=S,T$], the {\it effective} (decreased) lifetime of each component, is a function of the 
[N$_2$] concentration:
 \begin{equation}
 \frac{1}{\tau'_{j}}([N_{2}])~=~\frac{1}{\tau_{j}}~+~k~[N_2]
  \label{eq:tau_Qj}
 \end{equation}
and $A'_{j}$, the {\it effective} (quenched) amplitude,  is also function of the [N$_2$] concentration:
\begin{equation}
 A'_{j}([N_{2}])~=~\frac{A_j}{1+\tau_j~k~[N_2]}
 \label{eq:A_Qj}
 \end{equation}
Notice that under these definitions $\ell'(t)$ reduces exactly to $\ell(t)$ for [N$_2$]=0, namely for pure Argon.
The sum of the quenched amplitudes of Eq.~(\ref{eq:quench_pdf}) is no longer constrained to unity:
\begin{equation}
\int_0^\infty \ell'(t) dt~=~(A'_{S}~+A'_{T})~\le~1
\label{eq:sumAmpl}
\end{equation}

From Eq.~(\ref{eq:sumAmpl}) it can be defined the overall {\it quenching factor} ($Q_F$):
\begin{equation}
 Q_F~=~ A'_{S}~+~A'_{T}~~~;~~~0\le Q_F\le 1
 \label{eq:QF}
 \end{equation}
namely, the ratio between the total intensity of scintillation light emitted for a given [N$_2$] contamination 
with respect to the case of pure liquid Argon. $Q_F$ also represents by definition the normalization value for $\ell'(t)$.
 
This work is dedicated to a systematic study of the the effects of Nitrogen contamination in LAr. In particular, 
the aim is:
\begin{itemize}
\item to determine the overall quenching factor affecting the LAr scintillation light yield at various 
N$_2$ concentrations [Sect.~\ref{sec:SAA}];
\item to determine the effective lifetimes $\tau'_{j}$ as a function of the N$_2$ contamination for each component, as well as the quenched amplitudes $A'_{j}$ [Sect.~\ref{sec:SSA}];
\item to extract the main characteristics of the scintillation light emission in pure LAr and the value of the rate constant $k$ of the quenching process  associated to the presence of N$_2$ contaminant [Sect.\ref{sec:overall_fit}].
\end{itemize}

\section{Detector and experimental method}
\label{sec:LAr_Det}

A detector  has been designed and assembled at the INFN-LNGS external facility (``Hall di Montaggio'') for this specific 
task. It consists of a cylindrical LAr cell in PTFE (h=12 cm, $\varnothing$=8.5 cm internal dimensions, 
wall thickness 0.5 cm) containing about 0.7 lt of LAr (about 1 kg of active mass), viewed by a single 2" 
photo-multiplier ({\sf Electron Tubes ETL D745UA}) mounted on the top end of the  cell. \\
The photo-multiplier (PMT) is manufactured to work at cryogenics temperatures (LAr); however, the glass window is not transparent 
to the 127~nm LAr scintillation light. Therefore, the PMT optical window is covered by a layer of TetraPhenyl--Butadiene 
(TPB), an efficient frequency down-converter with emission spectrum  peaked at 430 nm and width of about 50 nm (FWHM), 
well matching the PMT sensitivity range. To enhance the light collection a reflector layer (also covered with TPB) 
surrounds the internal walls (side and bottom) of the PTFE cell.\\
\begin{figure}[htbp]
\begin{center}
\includegraphics*[width=10.7cm,angle=-90]{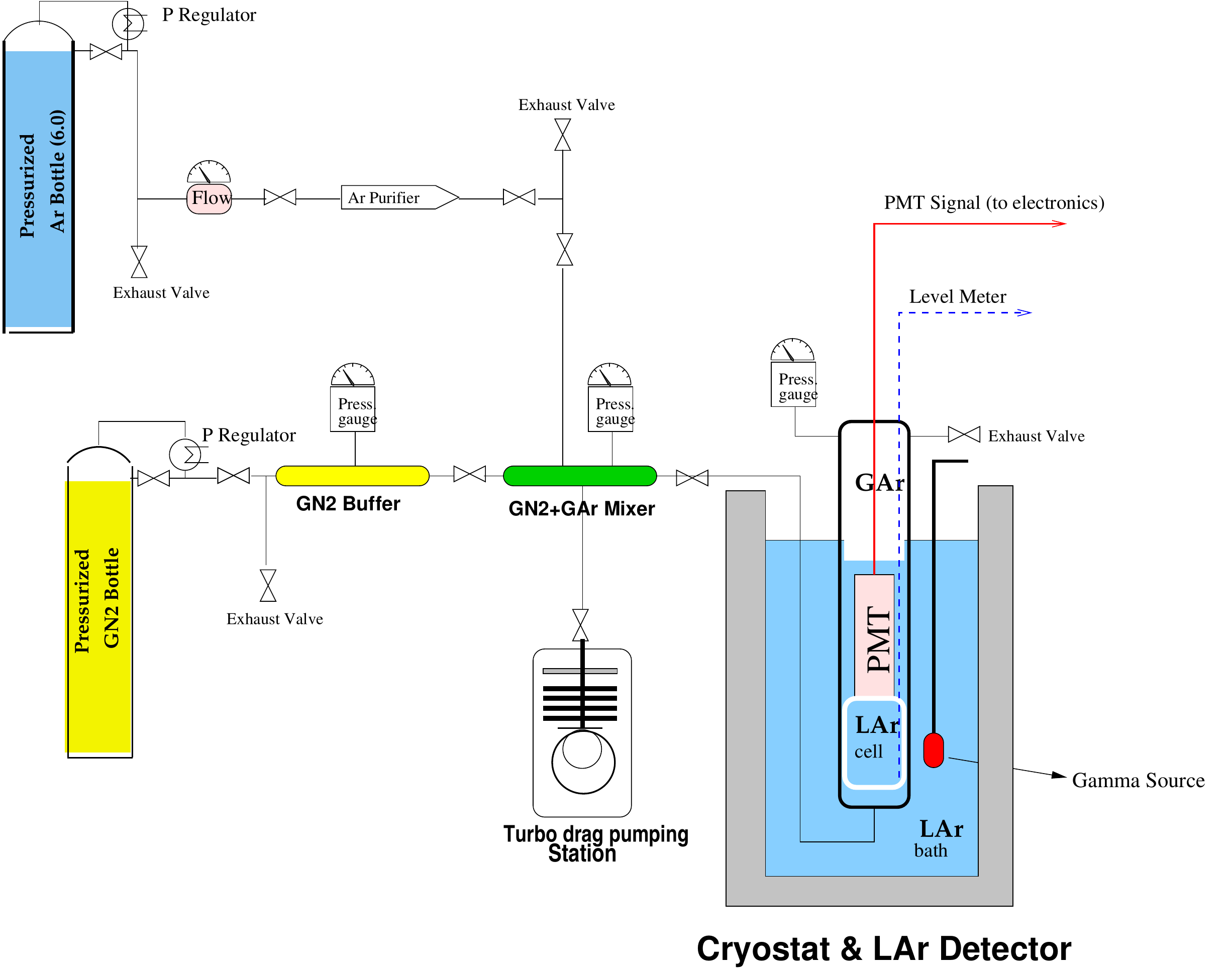}
 \caption{\textsf{\textit{Schematic layout of the experimental set-up for the N$_2$ contamination test.}}}
\label{fig:pmt}
\end{center}
\end{figure}
The detector is housed in a stainless steel cylindrical chamber (h=85 cm, $\varnothing$=10 cm), closed at both ends by 
vacuum-tight ConFlat (CF100) flanges, Fig.~\ref{fig:pmt} (schematic layout). The internal volume of the chamber is about 6.5 lt and it contains, after filling, 
a total amount of 3.0 lt of LAr (including the LAr cell active volume at its bottom).  The chamber is immersed in 
a LAr bath of a stainless steel open dewar, to liquify and keep at stable temperature the LAr internal volume.\\
A transfer line for the Ar filling and for the injection of controlled amounts of gaseous Nitrogen (GN$_2$) has been also assembled 
and connected to the LAr chamber via vacuum-tight pipes and feed-throughs. The line includes a two-stages system formed 
by a gaseous Nitrogen (GN$_2$) buffer and a mixer of gaseous Nitrogen and Argon, connected via UHV valves, Fig.~\ref{fig:pmt}. The internal 
volume of the buffer is precisely known and accurate pressure control allows the injection of controlled amounts of 
N$_2$ gas in the LAr volume (from 1 ppm to 300 ppm of N$_2$ per injection with $\le$ 6\% precision).\\
Gaseous Argon (GAr) for filling, supplied by a 200 atm pressurized bottle, is the best grade 6.0 (99.9999\%) commercial 
Argon, with impurity concentration below 1 ppm. In Tab.\ref{tab:Ar_60} the purity specification for the GAr in  use is reported \cite{rivoira}.
\begin{table}[htbp] 
\begin{center} 
\vspace*{0.2cm}
\caption{\textsf{\textit{Argon 6.0 (from Rivoira Supplier) purity specifications: maximum impurity levels.}}}
\vspace*{0.2cm}
\begin{tabular}{lr} 
\hline\hline 
Oxygen (O$_2$)             &        $\le$ 0.2 ppm        \\
Nitrogen (N$_2$)           &       $\le$ 0.5 ppm         \\
Water (H$_2$O)             &            0.5 ppm 	     \\
Hydrogen (H$_2$)         &            0.1 ppm       \\
Total Hydrocarbons (THC) &  	 0.05  ppm        \\
Carbon Dioxide (CO$_2$) 	&  0.05 ppm  \\
Carbon Monoxide (CO) 	&    0.05 ppm  \\
\hline\hline
\label{tab:Ar_60} 
\end{tabular} 
\end{center}
\end{table} 

\begin{figure}[htbp]
\begin{center}
\includegraphics*[width=10.0cm]{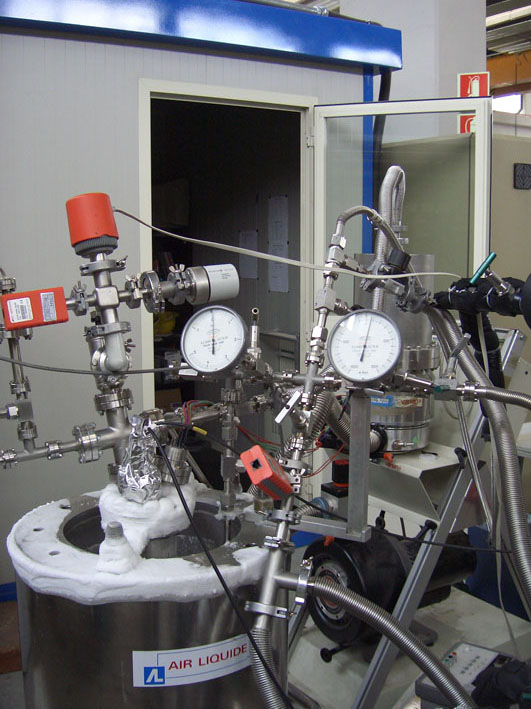}
 \caption{\textsf{\textit{Picture of the experimental set-up for the N$_2$ contamination test.}}}
\label{fig:pmt_photo}
\end{center}
\end{figure}

The GAr from the 6.0-bottle is flushed  during filling through a Hydrosorb/Oxysorb cartridge positioned 
along the GAr line for partial removal\footnote{No recirculation system is implemented for further O$_2$  purification. Initial content of residual N$_2$ in the 6.0 Ar is not filtered out. Additional contaminations from material out-gassing inside the chamber cannot be excluded.} of O$_2$ and H$_2$O. The GN$_2$ for the controlled contamination is 5.5 GN$_2$ type (99.9995\%), also supplied by a 
pressurized bottle.\\
In Fig.~\ref{fig:pmt_photo} a picture taken during the test is shown.\\
After vacuum pumping of the whole system (detector chamber and transfer line) up to a 
residual vacuum of few $10^{-5}$~mbar, the external bath is filled with 
commercial LAr, the GAr line is opened and the purified Ar gas feeding the detector chamber is liquified inside it 
up to reaching a given level, precisely monitored by a level meter, corresponding to 3.0 $\pm$1.5\% lt of LAr.

At LAr working conditions of T=86.5 K, P=930 mbar ($\rho_{Ar}$=1.40 g/cm$^3$), contamination with GN$_2$ is performed by a controlled procedure: (1) GN$_2$ corresponding to a given fraction of 
the LAr molar quantity\footnote{The concentration units in use here are ppm, {\it parts per million (atomic,} often indicated as ppma). This  indicates the ratio between the number of interesting elements (in our case N$_2$ molecules) to ordinary elements (Ar atoms).} in the chamber is first transferred into the GN$_2$ buffer volume (at controlled temperature and 
pressure),
(2) the valve to the GN$_2$+GAr mixer is opened, (3) GN$_2$ is injected into the chamber by repeated flushes of purified 
GAr through the mixer.\\
 Ar and N$_2$ form almost ideal vapor/liquid mixtures\footnote{The dilution process is rather slow. However, the speed is substantially increased by the repeated flushing of GN$_2$+GAr mixture through the liquid.}. In the liquid phase they are completely miscible and almost ideal 
liquid solution. The vapor pressure in the top part of the chamber (about 3.5 lt above the liquid level) is different for Argon and Nitrogen 
and can be determined by the Antoine equation assuming the gas at the same temperature of the liquid Ar. 
Nitrogen results to be more volatile than Argon by a factor of 3; therefore the Nitrogen concentration in gas phase is 
higher than in the liquid \cite{princeton}. However, since the gas/liquid volume ratio for Ar is very large (e.g.  835 at STP), the fraction 
of N$_2$ injected in the chamber which is diluted in the LAr volume is very close to one.

\subsection{Data taking}
\label{sec:data_tak}
\begin{figure}[htpd]
  \begin{center}
   \mbox{\epsfig{file=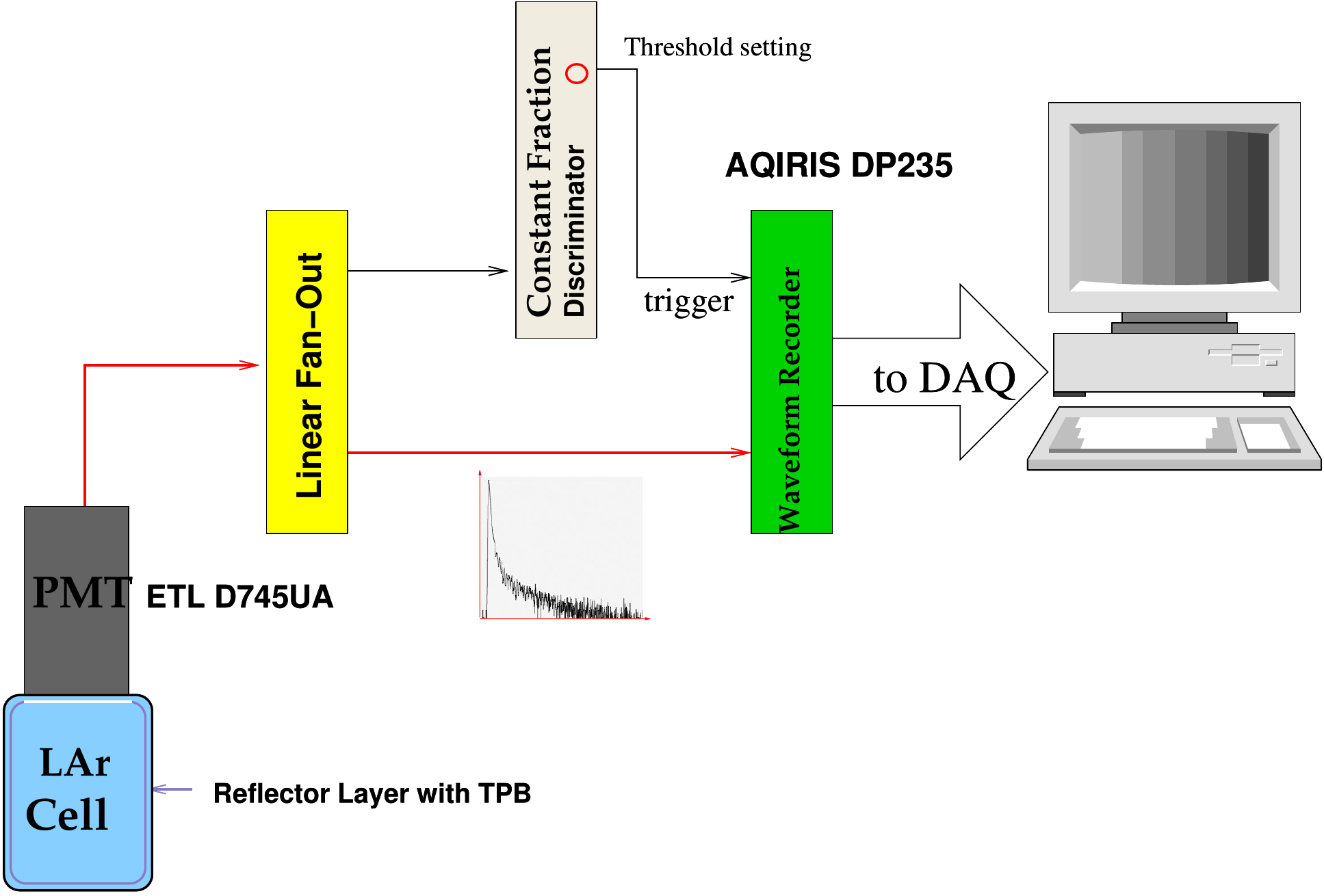,angle=-90,width=10.5cm}}
    \caption{\textsf{\textit{PMT Electronics read--out chain.}}}
  \label{fig:pmt_daq}
  \end{center}
\end{figure}

\begin{figure}[htbp]
\begin{center}
\includegraphics*[width=13cm]{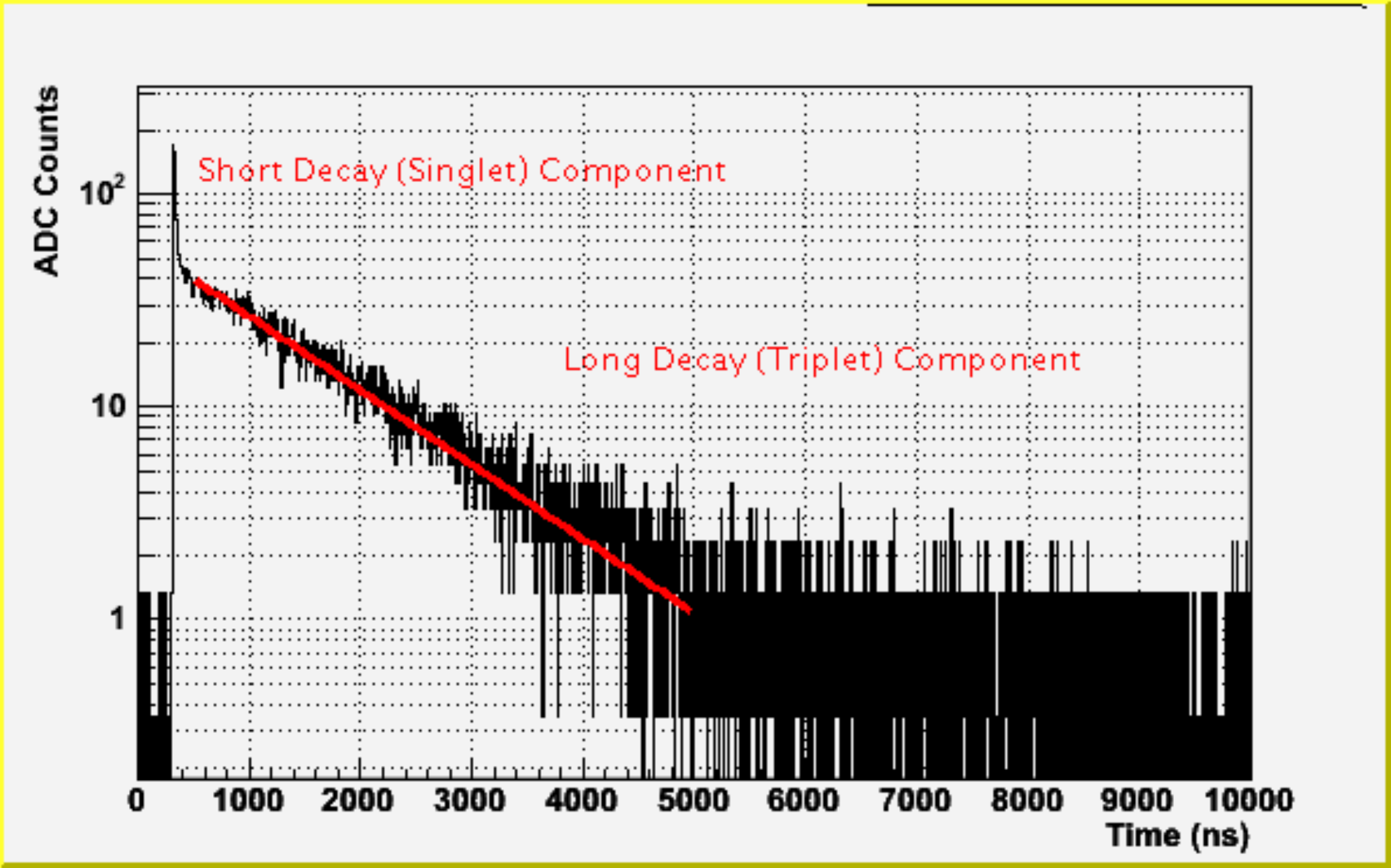}
 \caption{\textsf{\textit{Typical (single) waveform recorded during the N$_2$ test. 
 Event with large energy deposition from cosmic muon {(\mip)} crossing the LAr cell.}}}
\label{fig:wfm}
\end{center}
\end{figure}

The PMT read--out system is structured as an oscilloscope channel:
in fact, the PMT+read-out chain can be represented by the simplified scheme of Fig.~\ref{fig:pmt_daq},
where at the first stage the PMT signal is duplicated by a linear {\it Fan-out}, one output being used as input for a {\it Constant Fraction Discriminator} (CFD) providing a trigger signal when a defined threshold is passed, and the second output is directly recorded by a fast {\it Waveform Recorder} ({\sf Acqiris, DP235 Dual-Channel PCI Digitizer Card}). At each trigger from the CFD the signal waveform is recorded with sampling time of 1 ns  over a full 
record length of 10 $\mu$s. A {\sf LabView} application has been developed for the data acquisition and storage. 
In Fig.~\ref{fig:wfm} a typical waveform recorded during the N$_2$ test is shown.\\  
After the filling procedure with purified 6.0 Argon was completed, the N$_2$ experimental test started on February 2007 
and lasted about one month. The test was performed by adding progressively controlled amounts of N$_2$ and exposing the 
LAr cell to $\gamma$-sources after each contamination.
Data samples ($10^{5}$ waveforms per run, trigger rate $\sim$ 0.5 kHz) have been recorded in two independent runs per N$_2$ concentration, with 
$^{137}$Cs and $^{60}$Co $\gamma$-ray sources. After the first run at 0 ppm (i.e. no additional N$_2$ 
injection to the initial purified 6.0 Argon), the contamination levels were set at  [N$_2$] = 1 ppm, 
2 ppm, 7 ppm, 12 ppm, 20 ppm, 40 ppm, 60 ppm, 100 ppm, 500 ppm, 1000 ppm and 3000 ppm.\\
Single photo-electron measurements were also acquired before and after 
each $\gamma$-source run, to provide calibration data (see App.~\ref{sec:SER}) useful at various stages of the analysis performed. 

\vspace{-0.5cm}
\section{Signal Amplitude Analysis}
\label{sec:SAA}
\vspace{-0.5cm}
Gamma rays from the $^{60}$Co source (1.27 and 1.33~MeV) 
induce mainly single Compton interactions in the LAr cell active volume (Compton edge $E_C\simeq 1$ MeV), with final 
state electron in the {\it mip} range. Scintillation light is emitted and, after down-conversion and reflections 
at the active volume boundaries, is collected at the PMT photo-cathode; finally the signal waveform is recorded. \\
\begin{figure}[htbp]
\vspace{-.3cm}
\begin{center}
\includegraphics*[width=11cm]{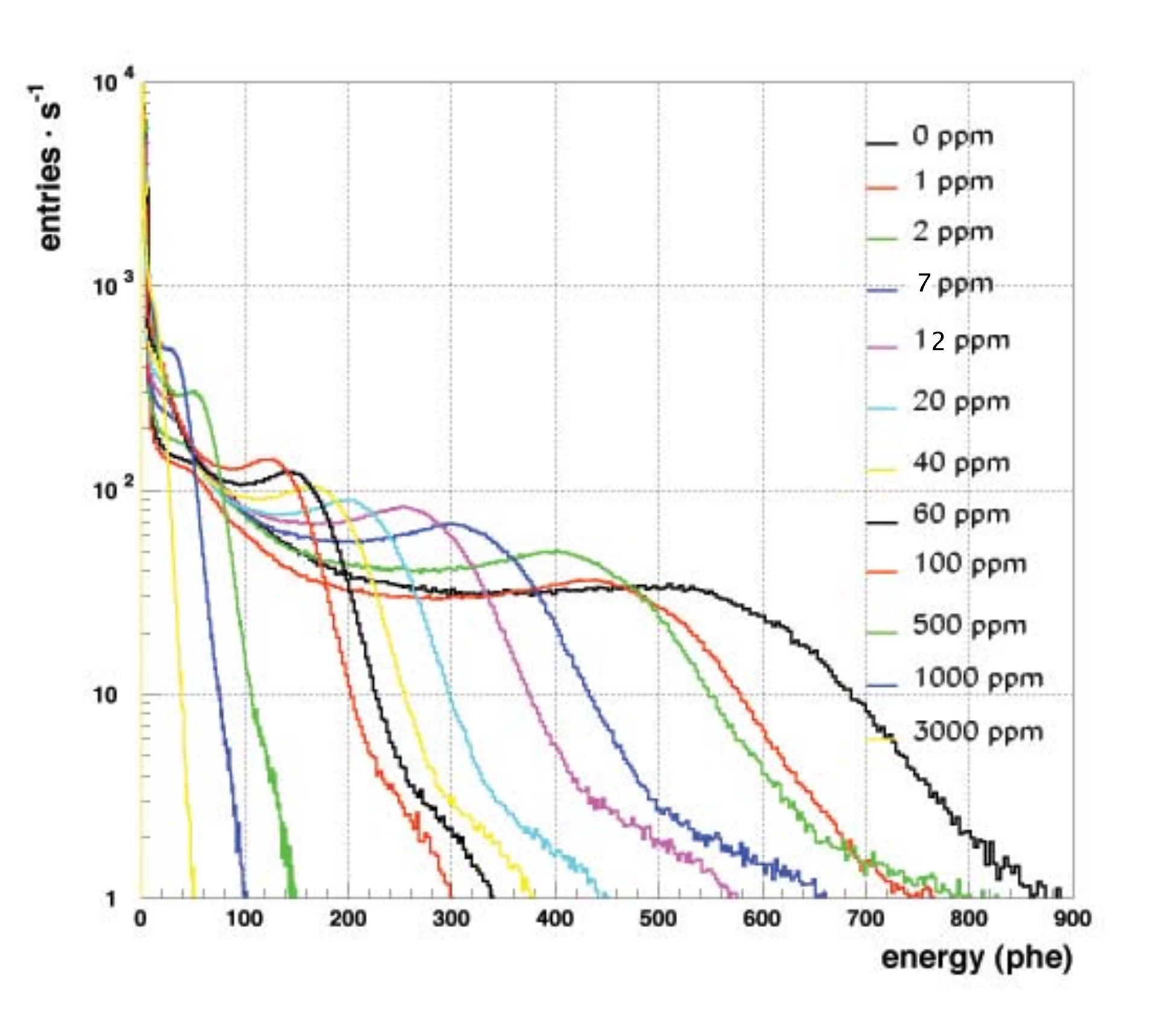}
\vspace{-0.8 cm}
 \caption{\textsf{\textit{$^{60}$Co source Compton spectra from each [N$_2$] run.}}}
\label{fig:Compton_spectra}
\end{center}
\end{figure}
By single waveform integration, the absolute (individual) signal amplitude is obtained (in units of 
photo-electrons, eventually proportional to the electron energy deposited in the LAr cell). \\
Pulse amplitude spectra have been thus obtained for each run at different [N$_2$] value, as reported in 
Fig.~\ref{fig:Compton_spectra} with the $^{60}$Co source. Manifestly, the 
spectra are progressively down-scaled at 
increasing N$_2$ concentration as due to the quenching process of reaction (\ref{eq:quench}).\\
\begin{figure}[htbp]
\begin{center}
\vspace{-.4cm}
\includegraphics*[width=8cm]{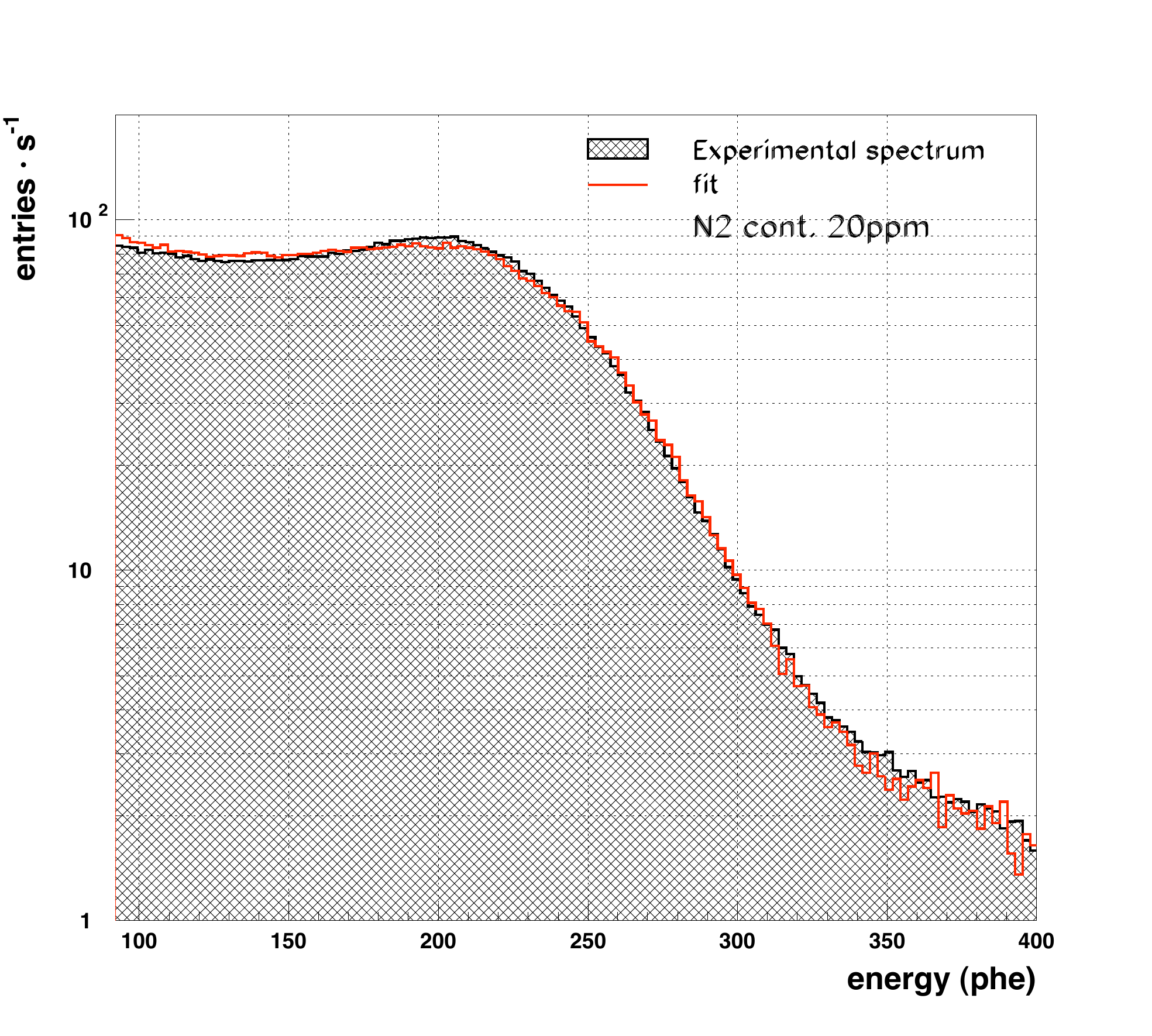}
\vspace{-.2cm}
 \caption{\textsf{\textit{$^{60}$Co source for the [N$_2$]=20 ppm run (dashed histogram). Superimposed the result of the fitting procedure (red histogram).}}}
\label{fig:Compton_fit}
\end{center}
\end{figure}
A dedicated fitting procedure has been developed to determine the scale factor Q$_F$ (the overall {\it Quenching Factor} defined in Eq.\ref{eq:QF}) acting on the absolute signal amplitude\footnote{Individual entry values of the uncontaminated (0 ppm) spectrum are down-scaled by Q$_F$, free parameter, to best fit the experimental [N$_2$] contaminated spectra.}
as a function of the N$_2$ contamination.  
As an example, in 
Fig.~\ref{fig:Compton_fit} the result from the fitting procedure for the [N$_2$]=20 ppm run is reported 
($Q_F$([20 ppm])=0.405: i.e. only $\sim$40\% of $Ar^*_2$ emit a photon, while the remaining $\sim$60\% undergo non-radiative N$_2$ collisions).\\

\begin{figure}[htbp]
\begin{center}
\vspace{-0.5cm}
\includegraphics*[width=13cm]{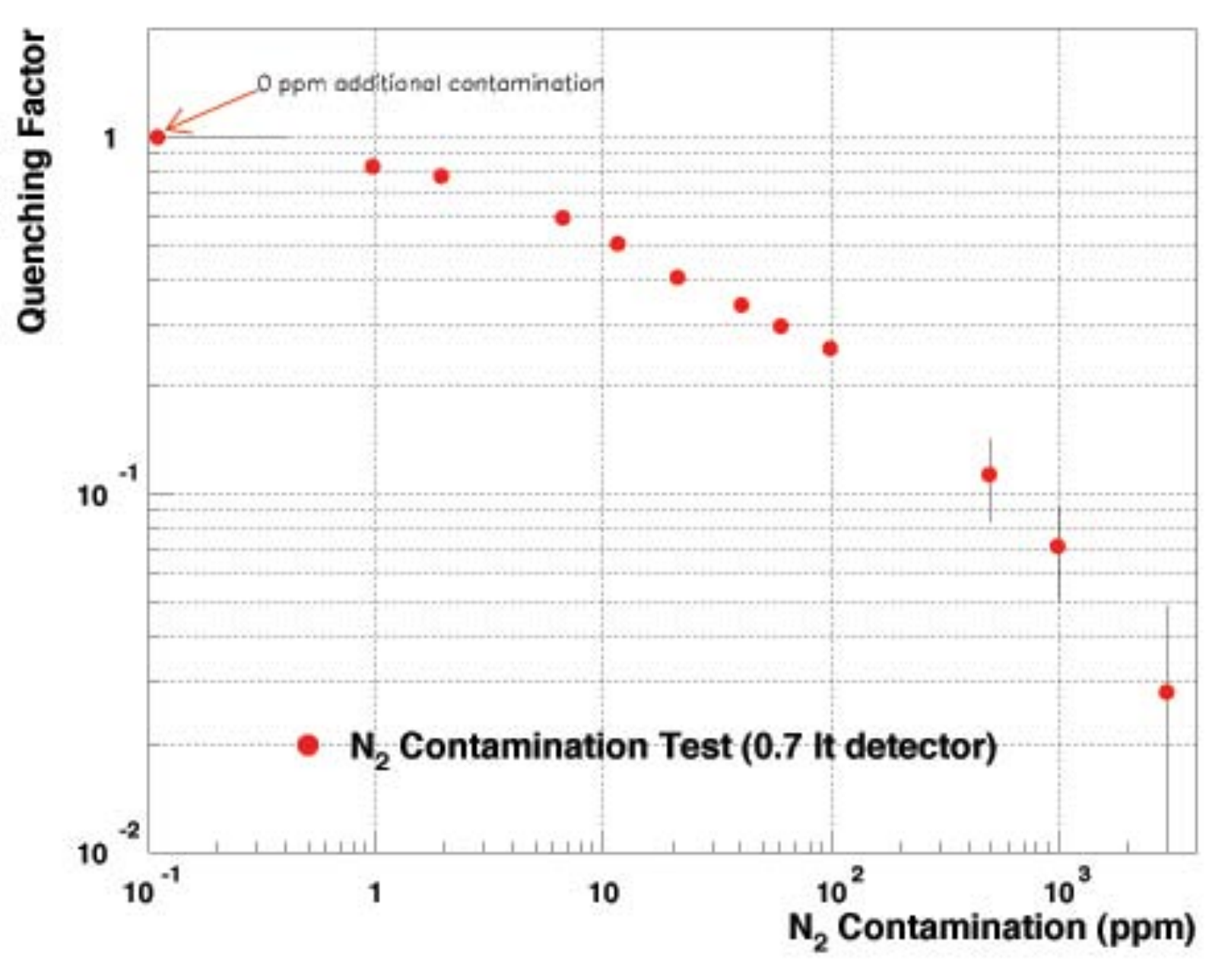}
 \caption{\textsf{\textit{Quenching factor Q$_F$ as a function of the [N$_2$] contamination ($^{60}$Co runs). The  value corresponds to the fraction of $Ar^*_2$ surviving N$_2$ quenching and producing VUV photons.}}}
\label{fig:QF_tot}
\end{center}
\end{figure}

The Q$_F$ values and associated errors obtained from the analysis of the whole set of $^{60}$Co runs at the various N$_2$ contaminations are given in Fig.\ref{fig:QF_tot}.
After a fast drop up to $\sim$ 100 ppm ($Q_F$([100 ppm])=0.25) a slower decrease is observed at higher N$_2$ 
concentrations (notice the {\it log-log scale} in use). The light contribution from the slow component  is expected to be vanished above [N$_2]\ge 100$ ppm, and the fast 
component starts to be progressively affected by the quenching process because of the very high Nitrogen concentration.\\

The errors on the plot are from the fit (statistical) and are appreciable at the highest concentrations.

\section{Signal Shape Analysis}
\label{sec:SSA}
\begin{figure}[htbp]
\begin{center}
\includegraphics*[width=12cm]{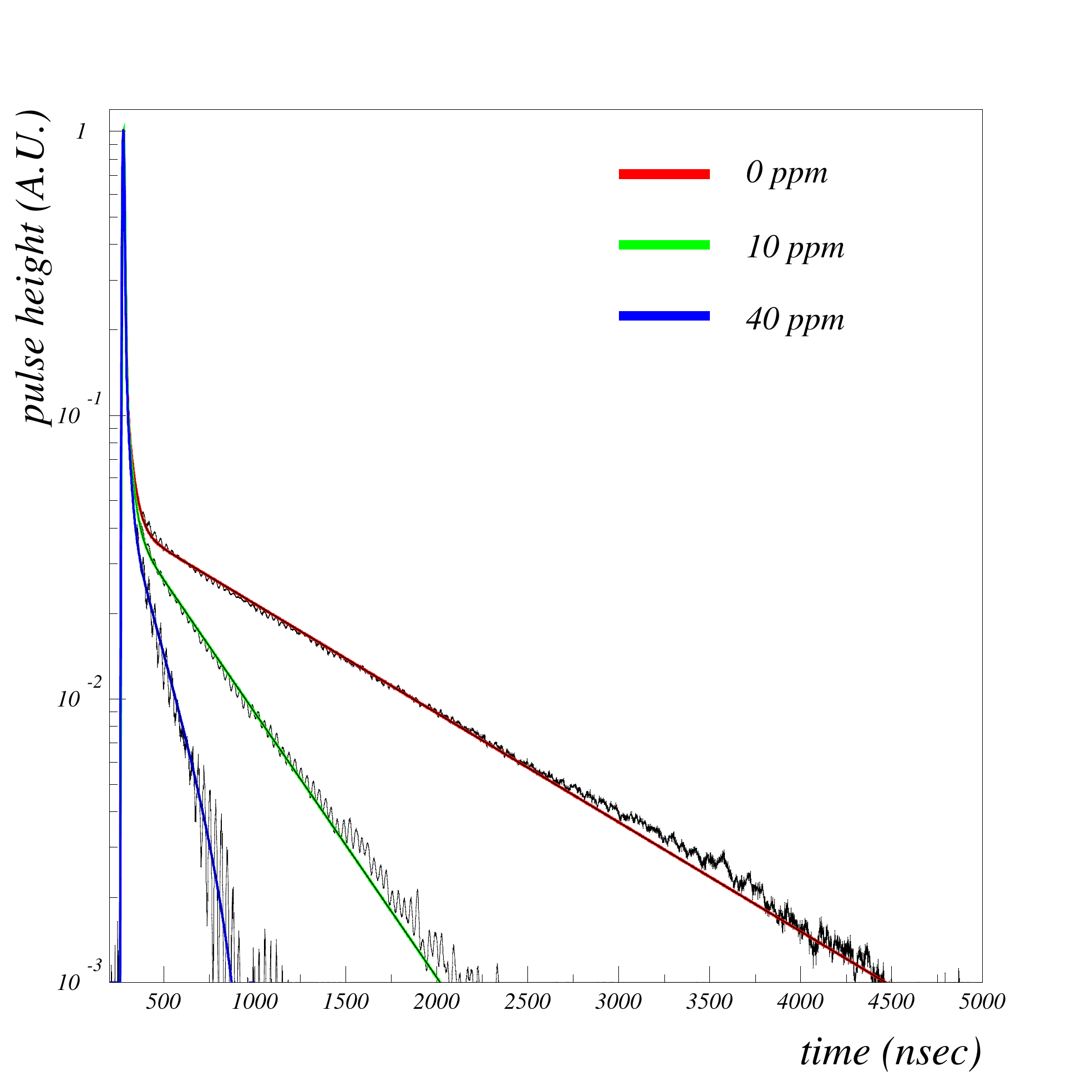}
 \caption{\textsf{\textit{Signal shape at 0 ppm, 12 ppm and 40 ppm of N$_2$ contamination, superimposed fits with three components.}}}
\label{fig:av_wfm}
\end{center}
\end{figure}

As reported in Sec.\ref{sec:LAr_Det}, data were collected by full signal waveform recording (1 ns {\it sampling time} over a {\it time window} of 10 $\mu$s) for each trigger event. This allows a detailed reconstruction of the signal shape, in particular of the individual components (relative amplitude and decay time) of the scintillation light following an ionization event in LAr.\\
To this purpose the recorded waveforms (wfm) have been appropriately  treated\footnote{Individual waveform (wfm) are off-line processed: (1) 0.5 $\mu$s of baseline presamples are used for baseline determination and subtraction, (2) cuts on wfm amplitude (integral) are applied to remove PMT saturations and low energy events, (3) wfm are time equalized at the same peak position and finally (4) selected wfm's are summed together.} 
and averaged ($\overline{V}(t)$ for each [N$_2$] run, 10$^5$ events) and  the averaged waveform amplitude (area) normalized to the corresponding Q$_F$ value.\\
The averaged signal can be symbolically expressed as:
\begin{equation}
\label{eq:convolution}
\overline{V}(t) = S(t) \otimes~\mathcal{R}(t)
\end{equation}
where $\mathcal{R}(t)$ is the {\it Response Function}  of the read-out system (PMT+cable+wfm digitizer), i.e. the electronic {\it impulse response} defined by the output signal obtained from an input single photo-electron pulse (usually indicated as SER, Single Electron Response).\\
The true light signal shape $S(t)$ can be obtained from the recorded waveform $\overline{V}(t)$ by a standard deconvolution procedure \cite{Num_Rec}, using the SER function $\mathcal{R}(t)$ experimentally obtained (as explained in App.~\ref{sec:SER}). In Fig.\ref{fig:av_wfm} three averaged signals (at [N$_2$]= 0 ppm, 12 ppm and 40 ppm) after SER deconvolution are shown.\\
The signal function $\ell'(t)$ (see Eq.\ref{eq:quench_pdf}) convoluted with a gaussian function has been used to fit the signal shape $S(t)$ from each [N$_2$] run. The gaussian spread (free parameter $\sigma_G\simeq 4$ ns from fit) takes into account the time resolution of the light collection at the detector sensitive area, due to transit time of the PMT, wave length shifting time, time spread due light propagation including multiple reflections at the volume boundaries, etc.\\
\begin{figure}[t]
\begin{center}
\vspace*{-2.5cm}
\includegraphics*[width=9cm]{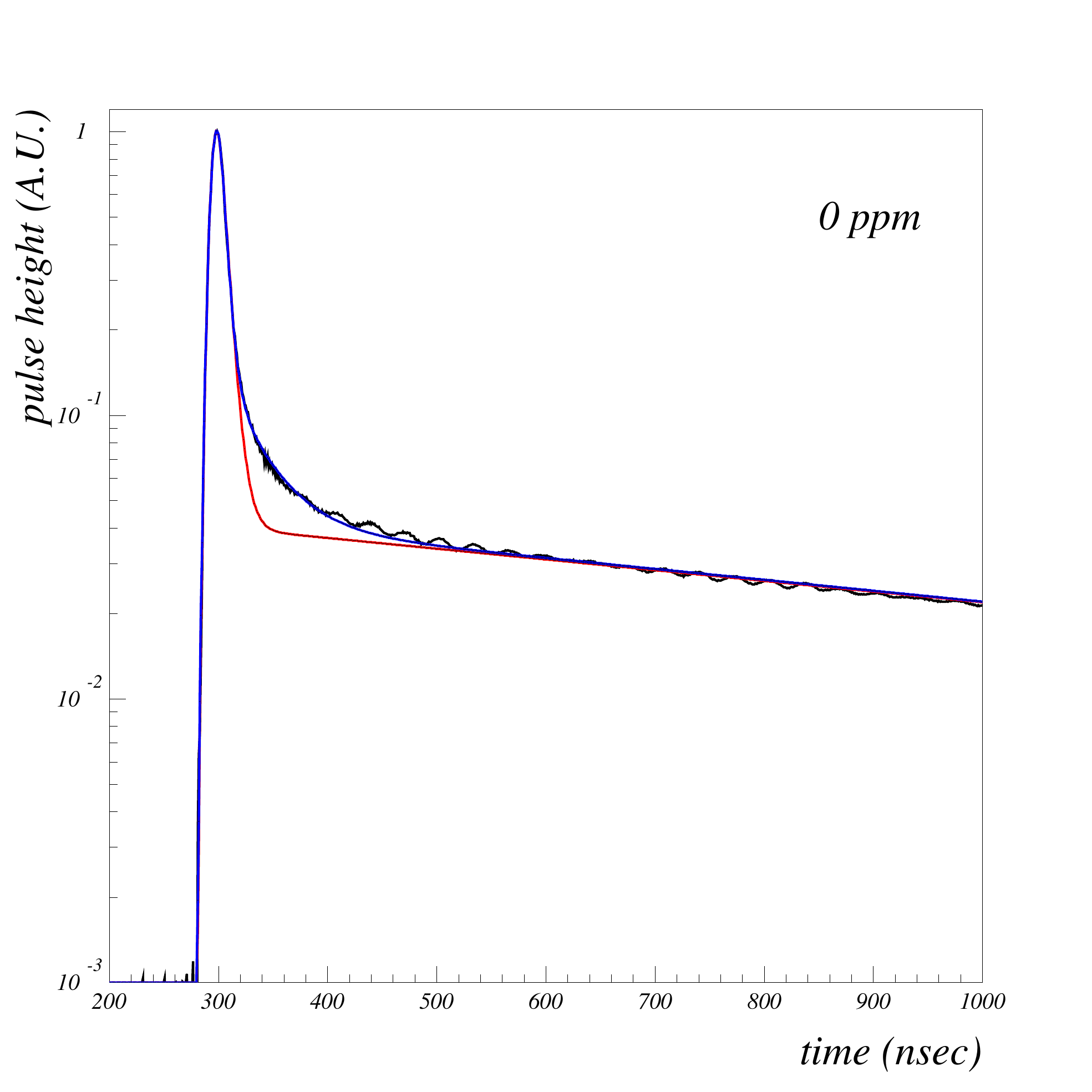}
 \caption{\textsf{\textit{Signal shape at 0 ppm contamination (black histogram) with comparison of two-components (red line) vs. three-component (blue line) model.}}}
\label{fig:two_vs_three_comp}
\end{center}
\end{figure}
Attempts to perform a fit with a signal function characterized by two time components (singlet and triplet) failed, as evident from Fig.\ref{fig:two_vs_three_comp}. This led to use a three components model (S, T and I, {\it Intermediate}), nicely fitting the present data (color lines in Fig.\ref{fig:av_wfm}).
The presence of an {\it intermediate} component is thus confirmed, in agreement with some earlier indications \cite{hitachi}. Dedicated studies on the origin of it are under way. Based on preliminary  considerations, reported in App.~\ref{app:int_comp}, possible instrumental effects appear as the main contribution to this component.\\
In Fig.\ref{fig:rel_ampl} the quenched amplitudes of the individual components  $A'_{j}$ from the fit are shown as function of N$_2$ contamination. 
The total amplitude (sum of the $A'_{j}$ amplitudes at given [N$_2$], black stars in Fig.\ref{fig:rel_ampl}) is constrained to the corresponding Q$_F$ value (Fig.\ref{fig:QF_tot}), as due to the average waveform normalization. \\
The decay times $\tau'_{j}$ of the individual components  from the fitting procedure of the signal shapes at different N$_2$ concentrations are reported in Fig.\ref{fig:tau_Q}. 

\subsection{Characteristics of LAr scintillation light and of N$_2$ quenching process}
\label{sec:overall_fit}
\begin{figure}[h]
\vspace*{-1.2cm}
\begin{center}
\includegraphics*[width=12.5cm]{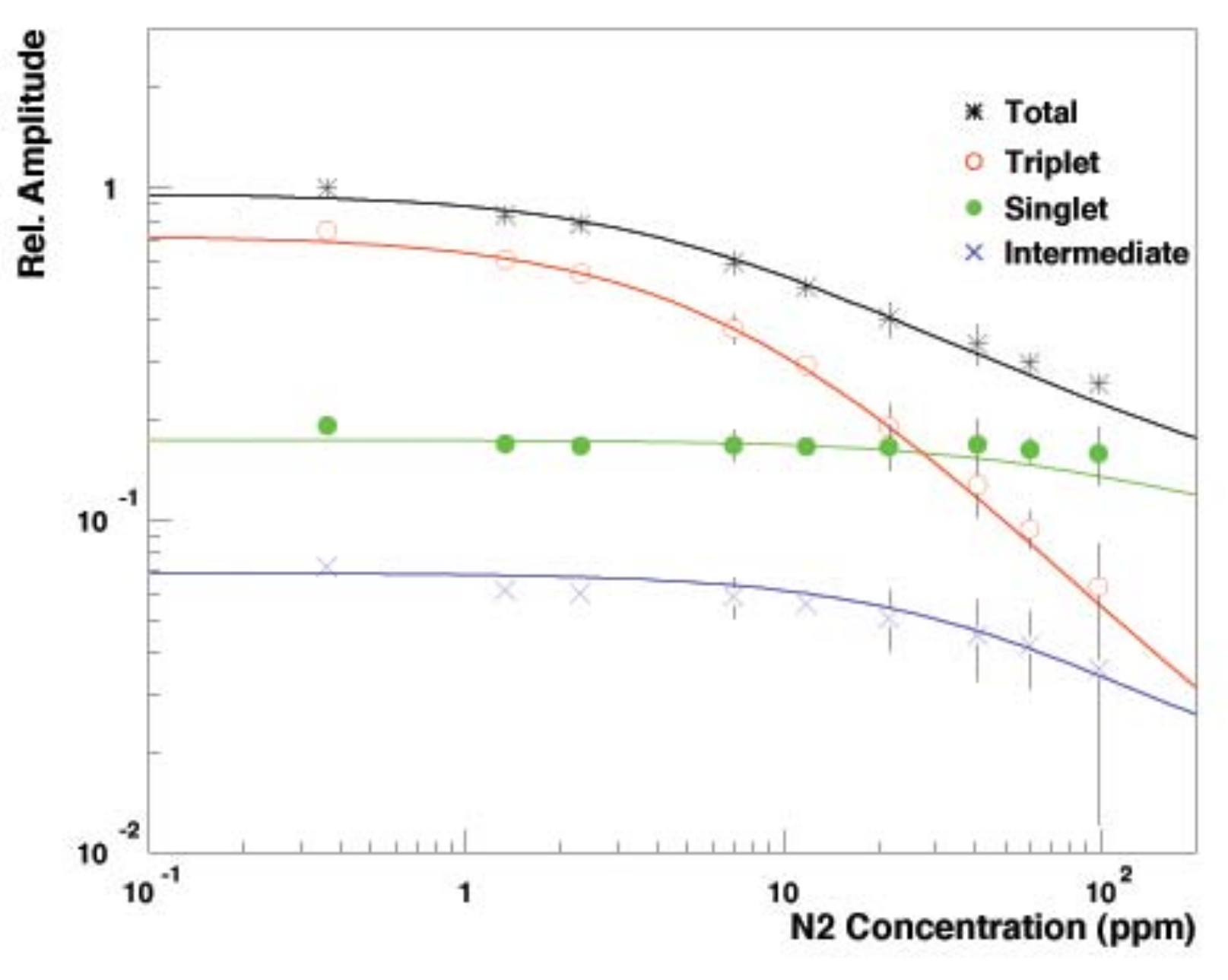}
\vspace*{-0.5cm}
 \caption{\textsf{\textit{Effective amplitudes ({\mip} electronic recoil) of the individual components 
 ($A'_{j},j=S,T,I$ in corresponding color an symbol) as function of N$_2$ concentration. 
 The total amplitude (sum of the $A'_{j}$ amplitudes at given [N$_2$]) is constrained to the corresponding Q$_F$ value (black stars) obtained in Sec.\ref{sec:SAA}. 
Lines in corresponding color are from the fit with saturation model (Eq.\ref{eq:A_sat}).
 }}}
\label{fig:rel_ampl}
\end{center}
\end{figure}

\begin{figure}[htbp]
\vspace*{-1.2cm}
\begin{center}
\includegraphics*[width=12.5cm]{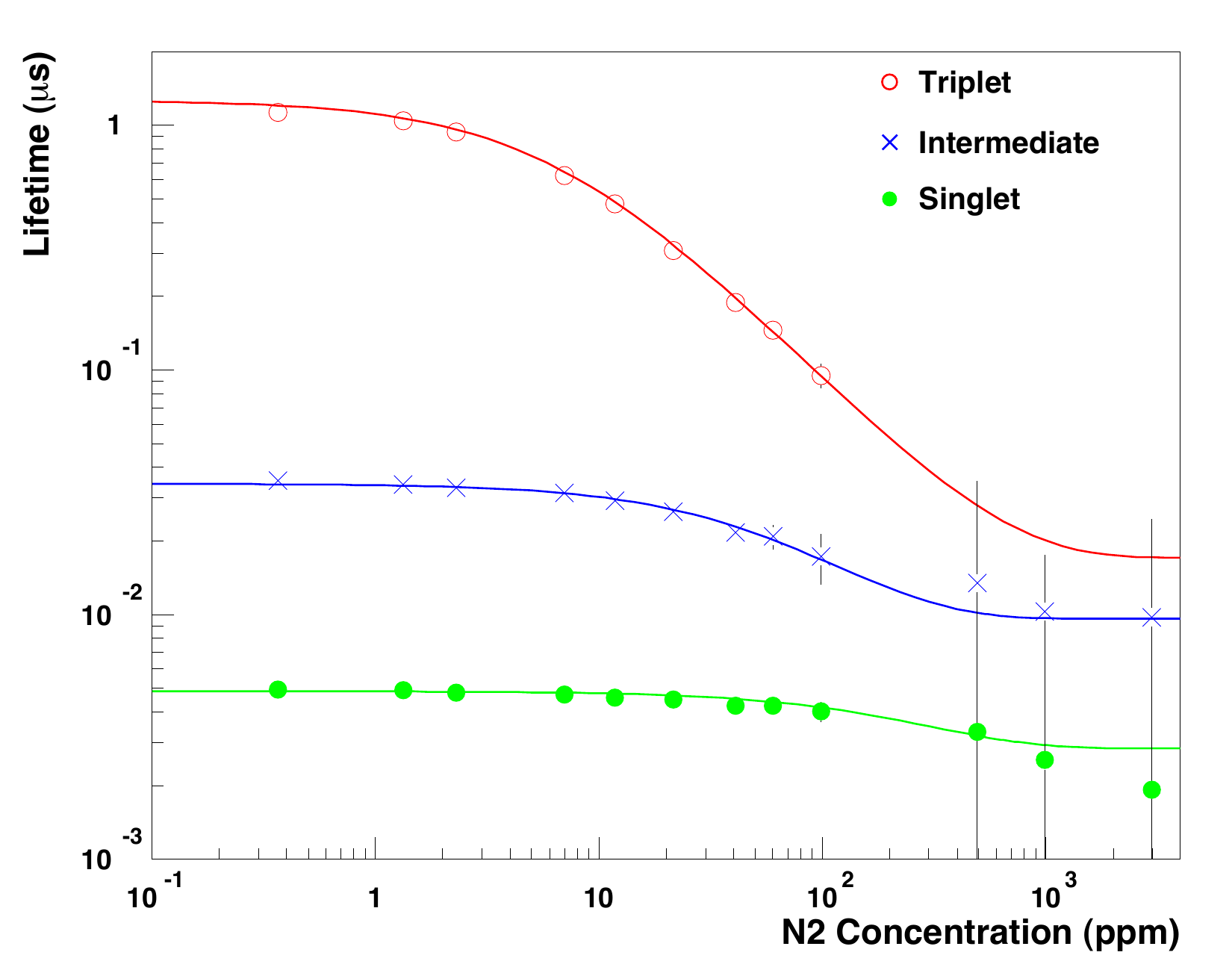}
\vspace*{-0.5cm}
 \caption{\textsf{\textit{Effective decay times of the individual components  ($\tau'_{j},j=S,T,I$ in corresponding color an symbol) as function of N$_2$ concentration. For [N$_2$]$\ge$500 ppm, $\tau'_T\rightarrow0$ (not shown). Lines in corresponding color are from the fit with saturation model (Eq.\ref{eq:tau_sat}).}}}
\label{fig:tau_Q}
\end{center}
\end{figure}

The quenched amplitudes and the decreased lifetimes of the scintillation light in N$_2$-contaminated LAr are correlated variables depending on the intrinsic LAr scintillation characteristics and on the 
rate constant of the N$_2$ quenching process, as reported in Sec.\ref {sec:light_quenching}.\\
An overall fit of the data obtained from the signal shape analysis reported in Fig.\ref{fig:rel_ampl} and Fig.\ref{fig:tau_Q} has been performed using Eq.\ref{eq:A_Qj} and Eq.\ref{eq:tau_Qj} as fitting functions, respectively.
The values of eight free parameters have been thus obtained by $\chi^2$-minimization: the LAr scintillation characteristics (intrinsic lifetimes $\tau_{j}$ and relative amplitudes $A_{j}$, $j=S,T,I$ for \mip~ electronic recoils), the rate constant $k$ of the quenching process due to N$_2$ contaminant, and the initial Nitrogen concentration [N$_2$]$_{in}$. This last corresponds to the N$_2$ content in the 6.0 grade Argon used for the initial filling. Its value is obtained by using an additive free parameter in the fitting functions: [N$_2$]$\rightarrow$[N$_2$]+[N$_2$]$_{in}$. \\
A contamination of Oxygen at the level of [O$_2$]=0.06 ppm has been taken into account into the fit, as determined by the analysis reported in \cite{O2_test} from the parallel test of Oxygen contamination effects. \\
The values of the parameters obtained from the overall fit are reported in Tab.2.\\
\begin{table}[htbp] 
\begin{center} 
\vspace*{0.2cm}
\caption{\textsf{\textit{Results from the overall fit: characteristics of scintillation radiation in LAr from {\mip} electronic recoil and of  quenching process from Nitrogen contamination. Quoted errors are statistical from the fit.}}}
\vspace*{0.2cm}
\begin{tabular}{ll|cr} 
\hline\hline 
Lifetimes            &        Amplitudes   & N$_2$ rate constant &   Initial N$_2$ concentration    \\
\hline
$\tau_S$~=~4.9$\pm$0.2 ns		&  	 $A_S$~=~18.8\%  &  			\\
$\tau_I$~=~34$\pm$3 ns		&  	 $A_I$~=~7.4\%    &  	$k~=~0.11\pm 0.01~$ppm$^{-1}~\mu s^{-1}$ & [N$_2$]$_{in}$~=~0.40$\pm$0.20 ppm		\\
$\tau_T$~=~1260$\pm$10 ns         &  	 $A_T$~=~73.8\%  &  			\\
\hline\hline
\end{tabular} 
\vspace*{0.2cm}
\end{center}
\label{tab:fit_param}
\end{table} 

The rate constant\footnote{The rate constant of the two-body (N$_2$-Ar$^*_2$) process is assumed to be the same for both the singlet $^1\Sigma_u$ and the triplet $^3\Sigma_u$ Ar excimer states. A fit has been tried with two rate constants as free parameters. This indicates that in case of 
N$_2$-$^1\Sigma_u$ coupling, a slightly higher constant rate is preferred.}
of the N$_2$ quenching process ($k$ value) is in agreement with an earlier measurement found in literature \cite{himi}.\\
The initial N$_2$ concentration is within the expected range from the specifications of the 6.0 grade Ar used for filling.\\
The relative amplitude of the intermediate component is small ($A_I$~=~7.4\%) compared to the amplitude of the other two components. It is also worth noticing that the amplitude ratio between 
$A_S+A_I$ and $A_T$ is 0.35 in agreement with available reference data for light ionizing particles \cite{hitachi}.

In all components the fit is found to deviate from the data at the highest N$_2$ concentrations, as possibly due to incomplete absorption of impurities into the liquid. This trend might be explained by a {\it saturation} effect of the solute (Nitrogen) in the LAr solvent. Similar indications have been reported by other groups \cite{himi} and also in our test with O$_2$ contamination \cite{O2_test}. The measurements were performed on relatively short time scale (hours to day per contamination). This  prevents to draw any definite conclusion on long term effects.\\
A different parameterization depending on the N$_2$ concentration under saturation hypothesis has been thus used for the fitting function of the lifetimes (each component):
 \begin{equation}
 \frac{1}{\tau'_{j}}~=~\frac{1}{\tau_{j}}~+~k~\beta~(1~-~e^{-\frac{[N_2]}{\beta}})
  \label{eq:tau_sat}
 \end{equation}
 and of the amplitudes:
 \begin{equation}
 A'_{j}~=~\frac{A_j}{1+\tau_j~k~\beta~(1~-~e^{-\frac{[N_2]}{\beta}})}
 \label{eq:A_sat}
 \end{equation}
where the $\beta$ parameter represents the concentration scale where saturation becomes effective ($\beta=530$ ppm from fit). At first order approximation both formulas reduce to the original fitting functions (Eq.\ref{eq:tau_Qj} and Eq.\ref{eq:A_Qj}).\\
The curves in Fig.\ref{fig:rel_ampl} and Fig.\ref{fig:tau_Q} (full line in corresponding color) show the result of the fit with the modified parameterization. This model improves the scaling with N$_2$ concentration and decreases the overall $\chi^2$, without changing at all the results reported in Tab.2.
 
\clearpage
The errors associated to the results in Tab.2 are from the fit (statistical). These take into account the statistical errors from the fit of the waveforms at different N$_2$ contaminations and also the error associated to the injected amounts of Nitrogen in the contamination procedure.\\
 The goodness of the overall fit (saturation model) is satisfactory (C.L.$\simeq$80\%).
Systematic uncertainty may come from different sources: instability of the signal read-out system, bias due to residual concentration of unknown types of impurities in LAr  leading to light absorption or further Ar-excimer quenching, possible uncertainty introduced by the procedures of data analysis.
Only this last issue has been investigated in some details. \\
The off-line analysis of the light signal shape (obtained from the PMT waveform recording at the different N$_2$ contaminations) is based on the deconvolution procedure of the SER function (see App.\ref{sec:SER}). 
Systematic uncertainty associated to this procedure is relevant mainly to the long-lived component determination and has been estimated to be around 4\%:   [$\tau_T~=~1260~\pm~10~(stat)~\pm~55~(syst)$ ns]. Comparisons with alternative fit procedures confirm this evaluation, as  reported in details in App.\ref{app:tau_t_discuss}.

\section{Conclusions}
New generation detectors based on LAr as active medium for Dark Matter direct search and for neutrino physics exploit both scintillation light and free electron charge from ionization events.\\
Residual content of electro-negative impurities in LAr (like Oxygen molecules) significantly reduces the amount of charge and light by attachment and quenching processes respectively. This led to the development of dedicated O$_2$ purification systems, routinely employed in the experimental set-up's.\\
 On the other hand, non electro-negative contaminants like Nitrogen can also be found at appreciable concentration level even in best grade commercial Argon.  Contamination at  ppm level of N$_2$ leads to a substantial reduction of the scintillation light intensity, due to the quenching process in two-body collision of N$_2$ impurities with Ar$^*_2$ excimer states otherwise decaying with VUV light emission.
 (No appreciable effects from N$_2$ contamination on the free electron charge can be expected, due to the low electron affinity of  the N$_2$ molecule).
 
A dedicated test has been performed by means of a controlled N$_2$ contamination procedure. 
Measurements have been done by \mip~ particle excitation (Compton electrons from $\gamma$-sources in the MeV range).\\ 
The effect on the scintillation light collection has been measured over a wide range of N$_2$ concentration, spanning from $\sim$10$^{-1}$ ppm up to $\sim$10$^{3}$ ppm, though a {\it saturation} effect of the solute (Nitrogen) in the LAr solvent has been presumably found at the highest contaminations. \\
The rate constant of the light quenching process induced by Nitrogen in LAr has been found to be $k$(N$_2$)=0.11 $\mu$s$^{-1}$ppm$^{-1}$ (in agreement with some early measurements reported in literature). This implies that, for example, at $\sim$ 1 ppm level a $\sim$ 20\% reduction of the scintillation light is experienced due to the N$_2$ quenching process. 
The quenching factor affecting the light yield has been determined over the whole explored range of N$_2$ concentrations.

The experimental test also allowed to extract the main characteristics of the scintillation light emission in pure LAr, thanks to the direct PMT signal acquisition by the fast Waveform Recorder in use with the implemented DAQ system.\\
The light signal shape is well represented by the superposition of three components with exponential decay. The fast and the slow components,
with decay time constants $\tau_S$=4.9 ns and $\tau_T$=1260 ns, are recognized to be associated to the decay of the singlet $^1\Sigma_u$ and the triplet $^3\Sigma_u$ excimer states of Ar$^*_2$. An intermediate component 
is also found to be present (as sometime reported in literature), whose origin could presumably be ascribed to PMT instrumental effects. 

The main effect of residual N$_2$ in LAr is of reducing the slow component lifetime at increasing concentration, and consequently of varying the ratio of the relative amplitudes to higher values.  This makes the slow signals from $\gamma/e$ background less distinguishable in Pulse Shape vs the fast Ar-recoils signals possibly induced by WIMP interactions, in Dark Matter search experiments. \\
Therefore, the implementation of dedicated methods for removal of the residual N$_2$ content results to be recommended with LAr based detectors, or at least the use of best grade commercial Argon with reduced Nitrogen contamination (below $\sim$0.5 ppm) appears as definitively necessary. 

\section{Acknowledgments}
\label{sec:Acknow}
We would like to warmly thank all the LNGS technical collaborators
which contributed to the construction of the detector and to its
operation.
In particular we acknowledge the contributions of E. Tatananni,
B. Romualdi and A. Corsi from the Mechanical Workshop, L. Ioannucci from the Chemical and Cryogenics Service, M. D'Incecco from the Electronics Workshop.\\
This work has been supported by {\it INFN} Istituto Nazionale di Fisica Nucleare, Italy), 
by {\it MIUR} (Ministero dell'Istruzione, dell'Universit\'a e della Ricerca, Italy) - Research Program Prot. 2005023073-003 (2005), 
by the {\it ILIAS} Integrating 
Activity (Contract RII3-CT-2004-506222) as part of the {\it EU FP6} programme in
Astroparticle Physics, by a grant of the President of the 
Polish Academy of Sciences and by {\it MNiSW} grant 1P03B04130.

\clearpage

\clearpage

\appendix
\section{Single Electron Response}
\label{sec:SER}
Single photo-electron pulses (SER data) from thermionic dark counts have been routinely acquired during the test, before and after each source run at the different N$_2$ concentrations.\\
\begin{figure}[h]
\vspace*{-0.5cm}
\begin{center}
\includegraphics*[width=9cm]{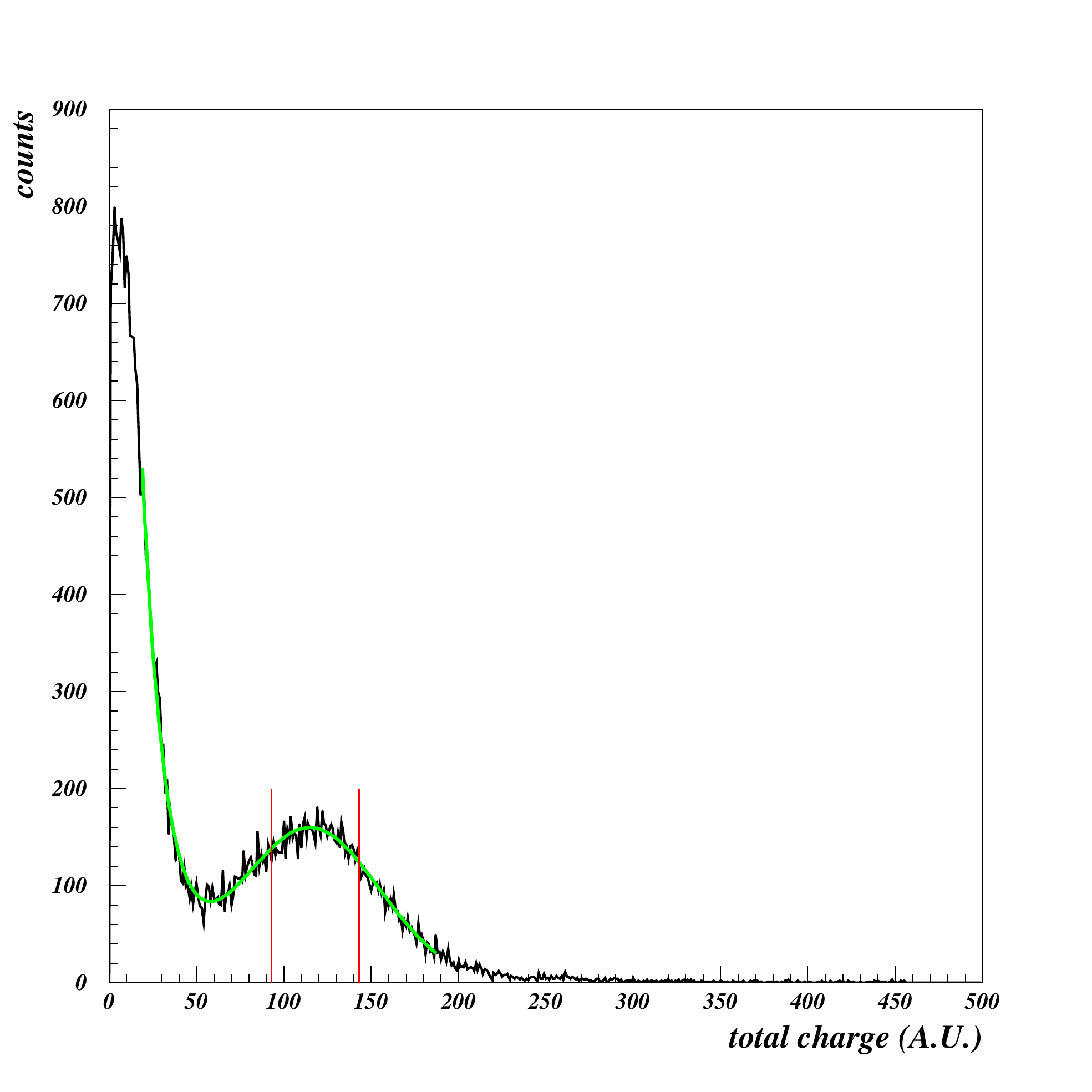}
 \caption{\textsf{\textit{SER spectrum of the PMT in use and fit superimposed. The interval around the mean value for single photo-electron waveform selection is also indicated.}}}
\label{fig:ser_spectrum}
\end{center}
\end{figure}
Pulse amplitude spectra from SER waveform integration were used for calibration purposes. Typical SER spectrum of the PMT in use is reported in Fig.\ref{fig:ser_spectrum}: the Gaussian distribution at higher ADC values corresponds to the genuine single photo-electron mean amplitude and spread.
The mean value (gaussian fit) gives the calibration constant, from ADC units into photoelectrons units, useful for Compton spectra analysis (Sec.\ref{sec:SAA}) and checking of the detector stability in time.

The single photo-electron pulses collected by full signal waveform recording (1 Gsample/s over 10 $\mu$s time window) were used for   
the determination of the read-out electronics  {\it Response Function}: the impulse response $\mathcal{R}(t)$ is obtained by processing and averaging the SER waveforms with amplitude in a defined interval around the mean value of  the corresponding single photo-electron pulse amplitude spectrum.\\
\begin{figure}[htbp]
\begin{center}
\vspace*{-0.5cm}
\includegraphics*[width=9cm]{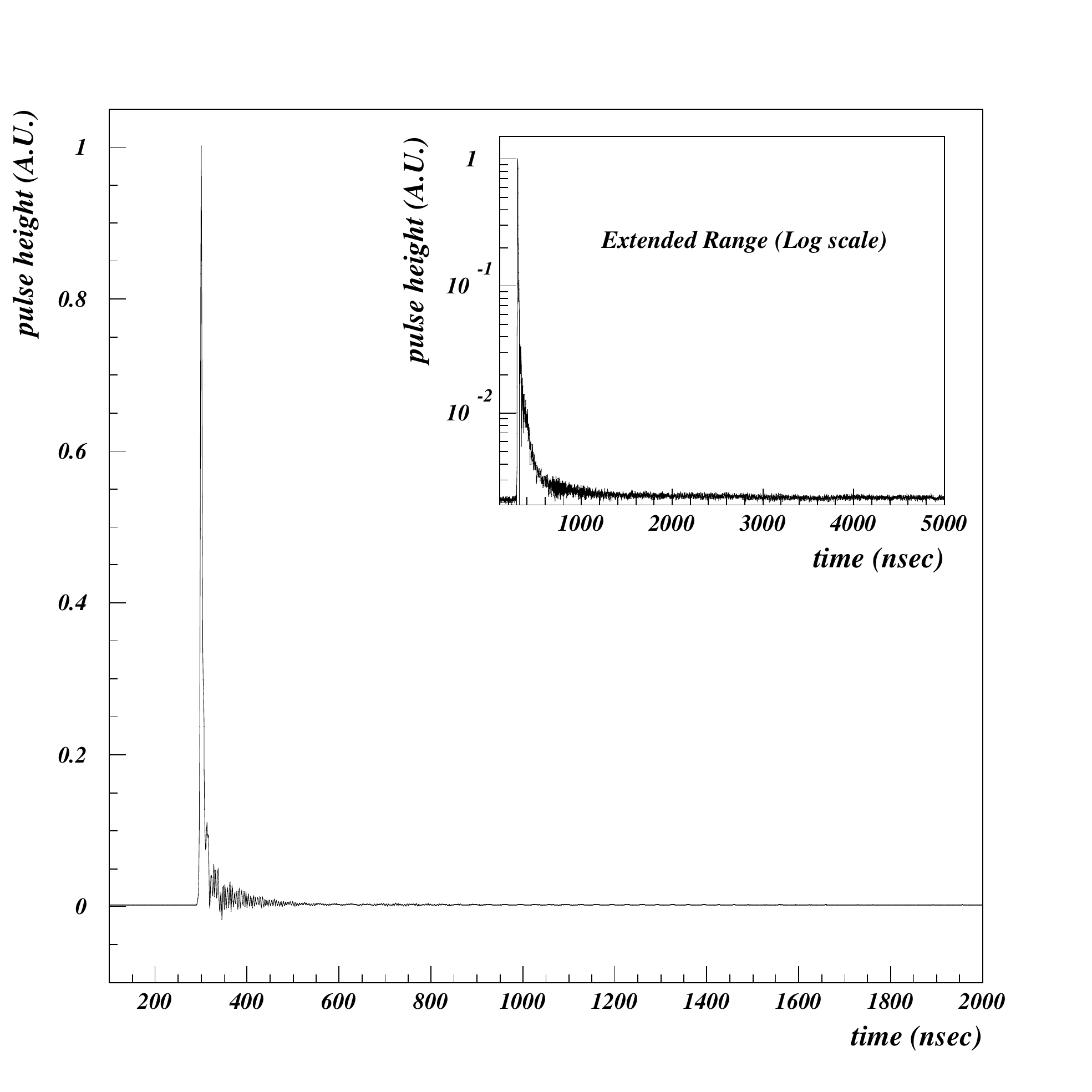}
 \caption{\textsf{\textit{Example of SER averaged waveform $\mathcal{R}(t)$.}}}
\label{fig:ser_wfm}
\end{center}
\end{figure}
A typical SER waveform averaged on a large sample of recorded signals is shown in Fig.\ref{fig:ser_wfm}. This is characterized by two main features: (1) the genuine single photo-electron shape from the PMT with FWHM $\simeq$ 10 ns and a decay tail extending up to about 150 ns, and underneath (2) a very long, slowly decreasing component up to $\sim 3\mu$s (visible in the up-right, log scale panel of Fig.\ref{fig:ser_wfm}) followed by a flat baseline.\\
 The flat baseline is due to pile-up (in the average wfm) from the PMT dark counts randomly distributed in the recorded time window, while the origin of the long-decay component is not fully understood. It may be ascribed to additional pile-up of {\it after-pulses} from ionization of residual gases in the PMT (normally reported in PMT test reports \cite{PMT_docs}). Typically, these type of after-pulses are delayed in a $\mathcal{O}$(0.5~-~5)~$\mu$s interval after the main signal from which they originate, in the ratio of $\le~5$\%  to the number of true pulses.

The SER function $\mathcal{R}(t)$ (including the pile-up components) determined for each source run is used in the deconvolution procedure to obtain the corresponding true light signal shape $S(t)$ (Sec.\ref{sec:SSA}).\\
The deconvolution of the response function $\mathcal{R}(t)$ from the experimental 
waveform $\overline V(t)$ is performed numerically in the frequency domain. The discrete 
Fourier transforms 
$\tilde{v}_i$ and $\tilde{r}_i$ of $\overline V_{i}$ and $\mathcal{R}_{i}$ respectively are first evaluated 
by the Fast Fourier Transform (FFT) algorithm~\cite{Num_Rec}. 
According to the convolution theorem, the true light signal $S_{i}$ is hence 
obtained by taking (with the same FFT algorithm) the inverse Fourier 
transform of $\tilde{s}_{i}$ (=~$\tilde{v}_{i}/\tilde{r}_{i}$).

\section{The intermediate component: preliminary considerations}
\label{app:int_comp}
As described in Sec.\ref{sec:SSA}, evidence for an intermediate component in the scintillation signal shape has been found at all the N$_2$ contaminations, including the ``0 ppm run" with lifetime $\tau_I\simeq$34 ns and relative amplitude $A_I\simeq$7.4\%. 
The origin of this component is controversial: it could be ascribed to (1) instrumental effects (in the PMT), (2) presence of residual N$_2$ concentration in LAr, (3) physical (unknown) properties of Argon.\\
Explanations in terms of:\\
- Intrinsic physical properties of Argon yet unknown or only marginally investigated cannot be excluded. For example, from  the delayed decay of vibrationally hot Ar$^*_2$ molecules surviving vibrational relaxation  \cite{koch}.  This decay, seen in gas at low pressure, was never investigated in liquid phase.\\
 - Light emission from Nitrogen trace in LAr is possible though rather unlike in the present experimental conditions:
 (a) scintillation emission from excited Nitrogen molecular states N$^*_2$(C$^3\Pi_u$) formed by {\it energy transfer} interaction with Ar$^*$ atoms, as seen in gaseous Ar-N$_2$ mixtures, should be negligible since in liquid phase the precursor Ar$^*$ atoms are quickly self-trapped\footnote{The Ar$^*$ self-trapping reaction rate is $\sim$10$^{12}$~s$^{-1}$ \cite{himi}, while the collision frequency of the energy transfer process is  $\sim$10$^{6}$~s$^{-1}$ at [N$_2$]=0.5 ppm (e.g. like in our ``0 ppm run").}, leading to Ar$^*_2$ formation
(moreover no increasing $A_I$ amplitude is found at increasing N$_2$ concentrations in Ar-N$_2$ studies [this paper] and \cite{himi}), however (b) a series of mechanisms leading to the formation of excited Nitrogen atomic states from Ar$^*_2$ energy transfer has been recently identified and detected in Ar-N$_2$(few \%) gas mixtures \cite{krilov}. The N$^*$ radiative decay (149 nm and 174 nm) may thus populate the intermediate component amplitude.  The size of this effect, presumably small in liquid mixtures with very low N$_2$ concentrations, as in the present study, is of interest for further studies.   \\
- Instrumental effects to explain the onset of the Intermediate component is 
our preferred option.
Such effects could be ascribed to the use of PMT and enhanced by particular experimental conditions. For example, the so called {\it anode glow} in the PMT \cite{PMT_docs} can be considered: it has long been known that electrons in PMTs induce light emission in the last cascades of
dynode systems. Photons may take the way back to the photo-cathode and so give rise to {\it after-pulses}, typically 20 to 50 ns after the true pulse from which they originate (and up to 120 ns in large PMTs).
This effect has been well characterized in various studies, indicating that the total amount is of the order of 1\%.\\
Contributions from {\it late-pulses} due to photo-electron
backscattering effect on the first dynodes can also be taken into account. Late-pulses are part of the main pulses  but delayed by 5 to 80 ns, in the ratio of $\le~5$\%  to the number of true pulses \cite{PMT_docs}. \\
Both contributions may  increase in high photo-electron density conditions, as in our case induced by the LAr fast decay component in the first few ns of scintillation emission.

All these are preliminary considerations. Only dedicated studies (outside the limits of this report) may possibly indicate the true origin of this component.

\vspace*{-0.5cm}
\section{Systematic error in the slow component lifetime determination}
\label{app:tau_t_discuss}
After completion of this report and while proof reading, 
an exhaustive paper on {\it ÒScintillation time dependence and pulse shape discriminationÓ} from Lippincott et al. 
\cite{McKinsey} appeared in publication. In this paper a value for the slow component lifetime  of $\tau_T$=(1463~$\pm$~5$_{stat}~\pm$~50$_{syst}$) ns is quoted. This value is only marginally compatible with the one we found here of $\tau_T$=(1260~$\pm$~10$_{stat}~\pm$~55$_{syst}$) ns. \\
We ascribe this discrepancy to the different strategies adopted in the analysis of  the LAr scintillation waveforms and not to physical or instrumental effects. In fact, by applying the procedure described in \cite{McKinsey} to fit the tail of the average waveform $\overline V(t)$ obtained from the ``0 ppm run" data sample we find a $\tau_T$ value for {\it pure} LAr in full agreement with the one reported in \cite{McKinsey}. \\
In more details, the value of  the slow component lifetime we obtain with the fit model  from \cite{McKinsey} (characterized by an additional flat baseline component)
\begin{equation}
\overline V(t)~=~A~exp(-t/\tau_T)~+B
\label{eq:fit_model_MK}
\end{equation}
 is $\tau_T$ =(1306~$\pm$~5) ns. Taking into account that the Argon we used to fill the cell (``0 ppm run") had indeed some residual contamination of Nitrogen and Oxygen (400 ppb of N$_2$ and  60 ppb of  O$_2$) and having determined the values of the quenching rate constants $k$ for N$_2$ [this report] and $k$'  for O$_2$ \cite{O2_test} respectively, we can extrapolate the $\tau_T$ value for pure LAr, that  results to be $\tau_T$=(1453~$\pm$~10) ns. \\
\begin{figure}[h]
\vspace*{-0.3cm}
\begin{center}
\includegraphics*[width=9cm]{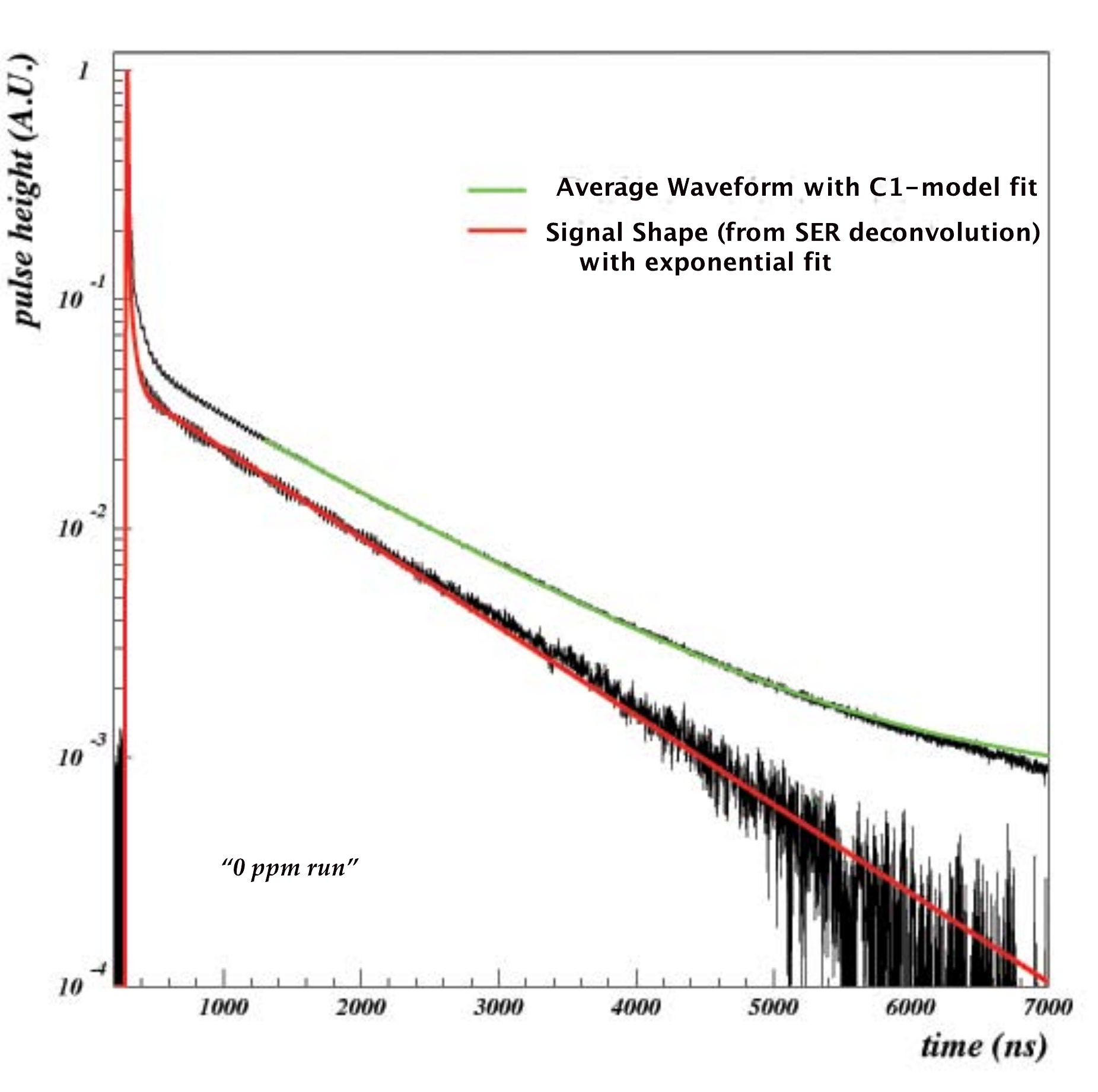}
\vspace*{-0.6cm}
 \caption{\textsf{\textit{Fit comparison: average waveform $\overline V(t)$, as built by summing up the waveforms recorded in the ``0 ppm run", with \ref{eq:fit_model_MK}-model fit (green line) and signal shape $S(t)$ (standard deconvolution procedure) with exponential fit (red line).}}}
\label{fig:tau_t_comp}
\end{center}
\end{figure}
 In Fig.\ref{fig:tau_t_comp} we show the average waveform $\overline V(t)$ with the fit (\ref{eq:fit_model_MK}) superimposed. For comparison the light signal shape $S(t)$ as obtained from $\overline V(t)$ with our standard procedure
 based on the SER function $\mathcal{R}(t)$ deconvolution is also shown.  The difference in the slope of the long-component is quite evident. In our data this is due to the presence of an ``extra-long tail" exponentially decaying with lifetime of $\sim$3 $\mu$s
 (rather than the simple flat baseline component), also present in the SER waveform (Fig.\ref{fig:ser_wfm}). 
 This possibly may fake a longer  triplet state decay time in the average non-deconvoluted waveform.
 Thanks to the deconvolution procedure this component is washed out.\\
It is worth noting that when alternatively a fit of $\overline V(t)$ with an extra-long decay component is performed ($B\rightarrow  B~exp(-t/\tau_B)$ in Eq.\ref{eq:fit_model_MK}, with $\tau_B\simeq 3~\mu$s) the triplet component lifetime is determined within 2\% spread around the value obtained from  our standard procedure (SER deconvolution).
 
 To further explore possible sources of systematic uncertainty due to the off-line procedures in our data analysis, we also tested the classic method of {\it Coincidence Single Photo-electron Counting} \cite{Bollinger} for another determination of the long component  lifetime.\\
The Coincidence Single Photo-electron Counting technique has been largely exploited in many reports on scintillation lifetime measurements (for LAr see \cite{hitachi},\cite{carvalho},\cite{himi}).
 Data recorded with our set-up (one single PMT with direct signal acquisition by fast Waveform Recorder) allow to perform an {\it artificial} single photo-electron counting experiment, based on off-line data treatment. 
 Starting at $0.5~\mu$s after the onset of a triggered signal, a single photo-electron identifying algorithm was run through the recorded waveform $V(t)$ and for each photo-electron pulse (defined by appropriate cuts) found at time $t_i$ a single bin with the value of 1 was substituted.\\
\begin{figure}[h]
\vspace*{-1.5cm}
\begin{center}
\includegraphics*[width=10.5cm]{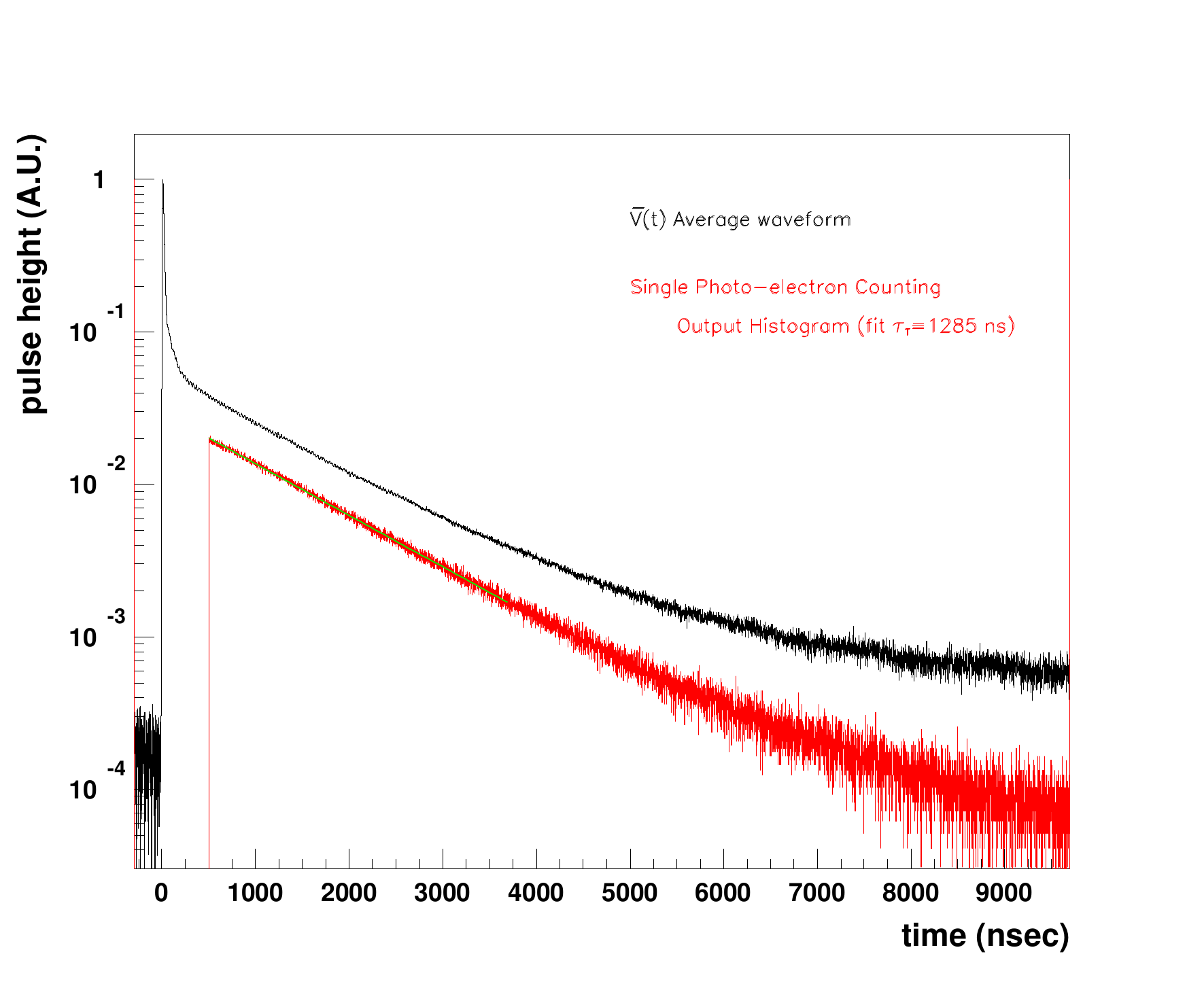}
\vspace*{-1.cm}
 \caption{\textsf{\textit{Result of the artificial single photo-electron counting method for $\tau_T$ determination (red histogram) with the ``0 ppm run" data sample. The average waveform $\overline V(t)$ (black histogram) is also shown for comparison.}}}
\label{fig:single_phel}
\end{center}
\end{figure}

  As an example with the ``0 ppm run" data sample, the resulting histogram, which is {\it per se} cleaned from noise contributions, provides a $\tau_{T}$ determination that is in very good agreement (within $\sim$~2\%) with the result of our standard procedure based on the SER function deconvolution,
 as shown in Fig.\ref{fig:single_phel}. 

\end{document}